\documentclass[11pt,a4paper]{article}
\pdfoutput=1
\usepackage{ifpdf}
\usepackage{jheppub}
\usepackage{rotating}
\usepackage[bb=boondox]{mathalfa}
\usepackage{comment}

\usepackage{tikz-cd}

\usepackage{graphicx} 
\usepackage{mathrsfs}
\usepackage{amsmath,amsfonts,amssymb}
\usepackage{mathtools}
\usepackage{graphicx}
\usepackage{dcolumn}
\usepackage{bm}
\usepackage{hyperref}
\usepackage[mathlines]{lineno}
\usepackage{float}
\usepackage{blindtext}
\usepackage{titlesec}
\title{Sections and Chapters}
\usepackage[toc,page]{appendix}
\usepackage{xcolor}
\usepackage{graphicx}
\graphicspath{ {./images/} }
\usepackage[thinc]{esdiff}
\usepackage[euler]{textgreek}
\usepackage{braket}
\usepackage{physics}
\usepackage{xfrac}
\usepackage{soul}
\usepackage{stmaryrd}
\usepackage{pifont}
\usepackage{caption}
\usepackage{makecell}
\usepackage{orcidlink}
\usepackage{soul}
\usepackage{multirow}

\allowdisplaybreaks

\title{\textcolor{black}{Electroweak Phase Transition, Gravitational Waves and Collider Probes in Multi-Scalar Dark Matter Scenarios}}

\author[a]{Tripurari Srivastava\,\orcidlink{0000-0001-6856-9517}\,}
\author[b,c]{\!\!, Jaydeb Das\,\orcidlink{0000-0001-6335-9377}\,}
\author[c]{\!\!, Anupam Ghosh\,\orcidlink{0000-0003-4163-4491}\,}
\author[d]{\!\!, Arnab Chaudhuri\,\orcidlink{0000-0002-6784-1360}\,}
\affiliation[a]{Institute of Particle Physics and Key Laboratory of Quark and Lepton Physics (MOE),
Central China Normal University, Wuhan, Hubei 430079, China}
\affiliation[b]{Indian Institute of Technology Kanpur, Kalyanpur, Kanpur 208016, Uttar Pradesh, India.}
\affiliation[c]{Department of Physics, Indian Institute of Technology Guwahati, North Guwahati, Assam-781039, India.}
\affiliation[d]{Division of Science, National Astronomical Observatory of Japan, Mitaka, Tokyo 181-8588, Japan.}
\emailAdd{tripurarisri022@gmail.com}
\emailAdd{jaydebphys@rnd.iitg.ac.in}
\emailAdd{anupamg@rnd.iitg.ac.in}
\emailAdd{arnab.chaudhuri@nao.ac.jp}

\abstract{
We study scalar singlet extensions of the Standard Model, focusing on scenarios where dark matter is stabilized by a \(\mathbb{Z}_2\) symmetry. In the minimal single-scalar extension of the SM, only a narrow region near the Higgs resonance remains viable, requiring small portal couplings in order to simultaneously satisfy the observed relic abundance and comply with the most recent direct detection limits from the LUX-ZEPLIN (LZ-2024) and XENON1T experiments. To address this limitation, we extend the dark sector by introducing additional real singlet scalars. In both two- and three-singlet extensions, we demonstrate that the observed dark matter relic density can be accommodated with larger Higgs portal couplings. These couplings significantly impact early-Universe dynamics by enhancing the strength of the electroweak phase transition. 
Both the two- and three-singlet scalar extensions can induce a strong first-order electroweak phase transition, generating stochastic gravitational waves potentially observable at future space-based detectors such as LISA and DECIGO. Notably, the three-singlet scenario induce an even stronger transition compared to the two-singlet case, enhancing the gravitational wave signal strength.
Our results highlight the potential of extended scalar sectors as testable frameworks connecting dark matter and gravitational wave signals.
}

\begin{document}
\maketitle
\flushbottom

\preprint{}

\section{Introduction}
\label{sec:intro}

The absence of a viable dark matter (DM) candidate within the Standard Model (SM) remains one of the most profound open questions in particle physics and cosmology~\cite{Clowe:2006eq,Sofue:2000jx}. Some of the simplest and most compelling extensions involve additional scalar fields stabilized by discrete symmetries. In particular, real scalar singlets and inert scalar doublets protected by a $\mathbb{Z}_2$ symmetry have long served as minimal and well-motivated DM candidates~\cite{Silveira:1985rk,McDonald:1993ex,Burgess:2000yq,Ma:2006km,LopezHonorez:2006gr,Deshpande:1977rw,Barbieri:2006dq,Ghosh:2021noq,Bhattacharya:2016ysw,Maity:2019hre,Ghosh:2025dcv}. However, the parameter space of these minimal models is now severely constrained by the increasingly stringent limits on the spin-independent DM–nucleon scattering cross section set by recent direct detection experiments, including XENON1T~\cite{XENON:2018voc} and LUX-ZEPLIN (LZ-2024)~\cite{LZ:2024zvo}.

An alternative approach involves extending the dark sector to include multiple real singlet scalars, each stabilized by an individual discrete symmetry. This leads to a multi-component dark matter scenario, where the total relic density is shared among several fields. In such frameworks, annihilation and semi-annihilation processes~\cite{Hambye:2009fg,Hambye:2008bq,DEramo:2012fou,Bhattacharya:2017fid} can contribute to relic abundance, with semi-annihilation often governed by couplings that do not directly mediate DM–nucleon scattering. As a result, these models can simultaneously satisfy relic abundance constraints and evade direct detection limits.

In this study, we consider an extension of the SM involving $n$ real scalar singlets, where each singlet is odd under its own discrete symmetry $\mathbb{Z}_2^{(i)}$. 
We explore a parameter space where the observed relic abundance is achieved via Higgs-resonant annihilation. The presence of multiple scalar dark matter candidates allows for larger portal couplings, while remaining consistent with direct detection constraints.
In addition to cosmological and astrophysical constraints, collider searches offer complementary and powerful probes of the Higgs portal interactions. In our scenario, the Higgs can decay invisibly into pairs of dark matter particles when kinematically allowed, providing a direct constraint from the Higgs invisible branching ratio measurements. Furthermore, monojet plus missing transverse momentum signatures at the Large Hadron Collider (LHC)—arising from dark matter pair production via an off- or on-shell Higgs boson also constrain the model. Together, these collider bounds can significantly restrict the viable parameter space of the model. In this work, we recast existing collider limits to assess their impact on our multi-scalar singlet framework.
We further examine whether the same parameter region can support a strong first-order electroweak phase transition (SFOEWPT), a feature of interest for probing the early universe. Finally, we investigate the associated gravitational wave signatures, which could be detected by future space-based interferometers. For theoretical constraints of our model such as vacuum stability and perturbative unitarity, we refer the reader to Refs.~\cite{Gonderinger:2009jp,Chao:2012mx}.

The possibility of a strong first-order electroweak phase transition is motivated not only by gravitational wave phenomenology, but also by scenarios of electroweak baryogenesis (EWBG)~\cite{Anderson:1991zb,Morrissey:2012db}. 
SFOEWPT is an essential ingredient in scenarios addressing the baryon asymmetry of the Universe. Specifically, it provides the necessary out-of-equilibrium dynamics, fulfilling one of the Sakharov conditions~\cite{Sakharov:1967dj}.  Although EWBG is not the focus of this work, its connection to phase transition dynamics further motivates the study of new physics scenarios capable of supporting a strong transition.

In the SM, the electroweak phase transition is known to be a crossover \cite{Kajantie:1996mn} for the observed Higgs mass of approximately 125 GeV~\cite{ATLAS:2012yve,CMS:2012qbp}, precluding the possibility of either EWBG or gravitational wave production from a thermal first-order transition~\cite{Kajantie:1996mn,Rummukainen:1998as,Csikor:1998eu}. A SFOEWPT can, however, be realized in extended scalar sectors by introducing additional bosonic degrees of freedom~\cite{Anderson:1991zb}, or via higher-dimensional operators~\cite{Zhang:1992fs,Camargo-Molina:2021zgz,Hashino:2022ghd,Oikonomou:2024jms,Gazi:2024boc}. Notable models include the Two-Higgs-Doublet Model (2HDM)~\cite{Cline:1996mga,Bhatnagar:2025jhh}, the (U)NMSSM~\cite{Apreda:2001us}, and various scalar extensions~\cite{Ellis:2022lft,Barger:2007im,Carena:2019une,Espinosa:2011ax,Noble:2007kk,Cline:2012hg,Alanne:2014bra,Cline:2013gha,Chaudhuri:2022sis,Chaudhuri:2017icn,Borah:2024emz,Chiang:2019oms,Kang:2017mkl,Kannike:2019mzk,Ghorbani:2024twk,Ghorbani:2019itr,DiazSaez:2024nrq,DiazSaez:2021pfw,Murai:2025hse}. In these frameworks, the strength of the transition can often be correlated with measurable quantities, such as deviations in the triple Higgs coupling~\cite{Kanemura:2004ch}.

A strong electroweak phase transition in the early Universe can produce a stochastic background of gravitational waves (GWs), providing an observational handle on the underlying scalar potential. The detection of GWs by aLIGO~\cite{Abbott:2016blz} has established gravitational wave astronomy as a new tool for probing early-universe dynamics. Cosmological phase transitions of first order can generate such stochastic GW backgrounds~\cite{Kamionkowski:1993fg}, which may be detectable by planned space-based interferometers such as LISA~\cite{eLISA:2013xep}, DECIGO~\cite{Kawamura:2011zz}, BBO~\cite{Corbin:2005ny}, as well as Chinese missions like TAIJI~\cite{Gong:2014mca} and TianQin~\cite{TianQin:2015yph}. The expected signal strength and frequency range of such GWs depend on the nature and dynamics of the phase transition. 

In this paper, we explore a simple framework involving multiple scalar singlet dark matter candidates and examine both their phenomenological and cosmological implications. We further analyze the nature of the electroweak phase transition within this model. Finally, we investigate the prospects of probing this scenario through the stochastic gravitational wave spectrum, with particular emphasis on its detectability by future interferometers.

%
%
%
%

The paper is organized as follows. In Section~\ref{sec:theory}, we outline the theoretical framework. Section~\ref{sec:dmanalyses} focuses on the dark matter phenomenology and compatibility with current experimental constraints. Constraints on the model parameters from the LHC data are discussed in Section \ref{sec:collider}. In the following Section~\ref{sec:ewpt}, we investigate the realization of a strong first-order electroweak phase transition within this framework by analyzing finite-temperature potentials. The associated stochastic gravitational wave spectrum is analyzed in the subsequent section. Finally, we present our summary and conclusions.

\section{Theoretical Framework}
\label{sec:theory}


We investigate a multicomponent scalar dark matter framework where an extended scalar sector plays a pivotal role in shaping the electroweak phase transition. The Standard Model is augmented with $n$ real gauge-singlet scalars $\phi_i$ $( i = 1, 2, \dots, n) $, each stabilized by an individual $ \mathbb{Z}_2^{(i)}$ symmetry, forming the group $\mathcal{G} = \mathbb{Z}_2^{(1)} \otimes \mathbb{Z}_2^{(2)} \otimes \cdots \otimes \mathbb{Z}_2^{(n)}$. SM fields are even under $\mathcal{G}$, while each $\phi_i$ is odd under its corresponding $\mathbb{Z}_2^{(i)}$ and even under the rest, ensuring the stability of each component. The resulting structure introduces rich dynamics in the thermal evolution of the early Universe, opening the possibility of a strong first-order electroweak phase transition.

\subsection{The Potential}
 The most general form of the renormalizable potential, which is invariant under $\mathcal{G}$ and the SM gauge group, can be written as
\begin{eqnarray}
   V(H,\phi_1,..\phi_n)  &=&   \mu_H^{2}|H|^{2} + \lambda_H |H|^{4} +\sum_i^n \left(\frac{\mu_{\phi_i}^2}{2}\phi_i^2  + \frac{\lambda_{\phi_i}}{4}\phi_i^4 + \frac{\lambda_{H\phi_i}}{2} |H|^2\phi_i^2\right) \nonumber \\
   &+& \sum_i^n\sum_{j>i}^n\frac{\lambda_{ij}}{4} \phi_i^2 \phi_j^2
  \, .
\end{eqnarray}
 Here, the SM Higgs doublet $H$ at $T=0$ is given by
\begin{equation}
   H= \frac{1}{\sqrt{2}}\begin{pmatrix}
        \chi_1 - i \chi_2\\
         h + v + i \chi_3
    \end{pmatrix}, 
\end{equation}
where $\chi_{1,2,3}$ are the Goldstone bosons and $v$ is the vacuum expectation value (\textit{vev}) of the Higgs field $h$ at zero temperature. After expanding the above potential with respect to classical background fields, the tree-level potential can be written in terms of classical background fields:
\begin{eqnarray}
    V_0(h,\phi_1,..\phi_n) & = &  \frac{\mu_H^{2}}{2} h^{2}+\frac{\lambda_H}{4} h^{4}+ \sum_i^n\left(\frac{\mu_{\phi_i}^2}{2}\phi_i^{2}+ \frac{\lambda_{\phi_i}}{4}\phi_i^{4} +  \frac{ \lambda_{H\phi_i}}{4} h^2\phi_i^2\right)
    + \sum_i^n\sum_{j>i}^n\frac{\lambda_{ij}}{4} \phi_i^2 \phi_j^2.\nonumber\\
\end{eqnarray}
Since every singlet scalar is a DM candidate, only the Higgs gets a vacuum expectation value ($vev$) at $T=0$, but others do not get $vev$. The following equation can be written from the minimization condition of the tree-level potential and Higgs mass $m_h^2$, as there is no mixing between the Higgs field and any of the singlet scalar fields. 
\begin{align}
\frac{\partial V_0}{\partial h}\Big|_{h=v,\phi_i=0} =0, \quad  m_h^2 = \frac{\partial^2 V_0}{\partial h^2}\Big|_{h=v,\phi_i=0} \longrightarrow \mu_H^2 = -\frac{1}{2} m_h^2, \quad
\lambda_H =  \frac{m_h^2}{2 v^2}.
\end{align}
Here, $v(\approx 246~\text{GeV})$ is the $vev$ of the SM Higgs field at $T=0$ as mentioned earlier, and $m_h = 125$ GeV is the mass of the Higgs boson. The bare mass parameter $\mu_{\phi_i}^2$ for every singlet scalar field can be written in terms of their physical masses as
\begin{eqnarray}
  \mu_{\phi_i}^2 = m_{\phi_i}^2-\frac{1}{2}\lambda_{H\phi_i}v^2.
\end{eqnarray}
 Note that all singlet scalar masses are the input parameters, as well as Higgs portal couplings $\lambda_{H\phi_i}$, scalar quartic couplings $\lambda_{\phi_i}$, and self-interacting couplings $\lambda_{\phi_i\phi_j}$ are free parameters here.

\section{Dark Matter Candidate and Relic Abundance}
\label{sec:dmanalyses}
Cosmological observations, such as those by the Planck satellite, indicate that approximately \(26\%\) of the total energy density of the Universe consists of non-baryonic dark matter (DM). While gravitational evidence for DM is well-established, its particle nature remains one of the most profound open questions in fundamental physics. Theoretical models typically classify DM production mechanisms into thermal and non-thermal categories.

In this study, we explore a scalar dark matter candidate stabilized by a discrete \(\mathbb{Z}_2\) symmetry. Within this setup, DM undergoes a thermal freeze-out: initially in equilibrium with the Standard Model (SM) plasma via Higgs-portal interactions, the DM particles decouple as the Universe expands and cools. The resulting relic abundance is governed by the annihilation process \( \phi_i \phi_i \rightarrow \text{SM SM} \), where \(\phi_i\) denotes a real singlet scalar.

We assume all singlet scalars \(\phi_i\) couple to the SM Higgs with similar strength, and that only the Higgs field acquires a vacuum expectation value ($vev$), while the \(\phi_i\) fields remain inert with vanishing $vev$s at zero temperature. All calculations are performed using \texttt{MicrOMEGAs}~\cite{Belanger:2006is}, which solves the relevant Boltzmann equations.

\subsection*{Relic Density Computation}

The evolution of the dark matter number density in the early Universe is governed by the Boltzmann equation. For a scalar dark matter component \(\phi_1\), the equation takes the form:
\begin{eqnarray}
    \frac{dY_{\phi_1}}{dx} &=& -\frac{x\,s(x)}{H}\, \langle \sigma v \rangle_1 \left(Y_{\phi_1}^2 - Y_{eq}^2 \right),
\end{eqnarray}
where \(x = \frac{m_{\phi_1}}{T}\), \(Y_{\phi_1} = n_{\phi_1}/s\) is the comoving number density, and \(Y_{eq}\) is its equilibrium counterpart. The entropy density \(s(x)\) is given by:
\begin{eqnarray}
    s(x) &=& \frac{2\pi^2}{45} g_* \frac{m_{\phi_1}^3}{x^3},
\end{eqnarray}
with \(g_*\) representing the effective number of relativistic degrees of freedom, and \(H\) being the Hubble expansion rate. The thermal average \(\langle \sigma v \rangle_1\) denotes the annihilation cross-section averaged over the thermal distribution of particles and is expressed as:
\begin{eqnarray}
    \langle \sigma v \rangle_1 &=& \frac{1}{n_{eq}^2} \frac{m_{\phi_1}}{64 \pi^4 x} \int_{4m_{\phi_1}^2}^{\infty} \hat{\sigma}(s) \sqrt{s} K_1 \left( \frac{x \sqrt{s}}{m_{\phi_1}} \right) ds,
\end{eqnarray}
where the equilibrium number density \(n_{eq}\) is
\begin{eqnarray}
    n_{eq} &=& \frac{g_{\phi_1}}{2\pi^2} \frac{m_{\phi_1}^3}{x} K_2(x),
\end{eqnarray}
and the reduced cross-section \(\hat{\sigma}(s)\) includes the standard phase-space suppression:
\begin{eqnarray}
    \hat{\sigma}(s) &=& \hat{\sigma}_0 \cdot g_{\phi_1}^2 \cdot \sqrt{1 - \frac{4m_{\phi_1}^2}{s}}.
\end{eqnarray}
Here, \(g_{\phi_1} = 1\) for a real scalar, and \(K_1\), \(K_2\) are modified Bessel functions of the second kind.

The relic abundance is finally calculated as:
\begin{eqnarray}
    \Omega_{\phi_1} h^2 &=& 2.742 \times 10^8 \left( \frac{m_{\phi_1}}{\text{GeV}} \right) Y_{\phi_1}(\infty),
\end{eqnarray}
where \(Y_{\phi_1}(\infty)\) is the asymptotic value of the yield after freeze-out. This expression includes standard cosmological constants for entropy and critical density. All numerical results presented in the following are generated using \texttt{MicrOMEGAs}.

\subsection*{Dark Matter Evolution: scalar singlet status}
We begin by exploring the parameter space that satisfies both the observed dark matter relic abundance and current bounds from direct detection experiments such as LUX-ZEPLIN (LZ)~\cite{LZ:2022lsv} and XENON1T~\cite{XENON:2018voc}. In this analysis, we consider a single real scalar singlet, stabilized by a \(\mathbb{Z}_2\) symmetry, as the dominant dark matter candidate. The relic abundance and spin-independent DM–nucleon scattering cross-section are computed using the \texttt{MicrOMEGAs} package.

\begin{figure}[h]
    \centering
    \includegraphics[width=0.48\textwidth]{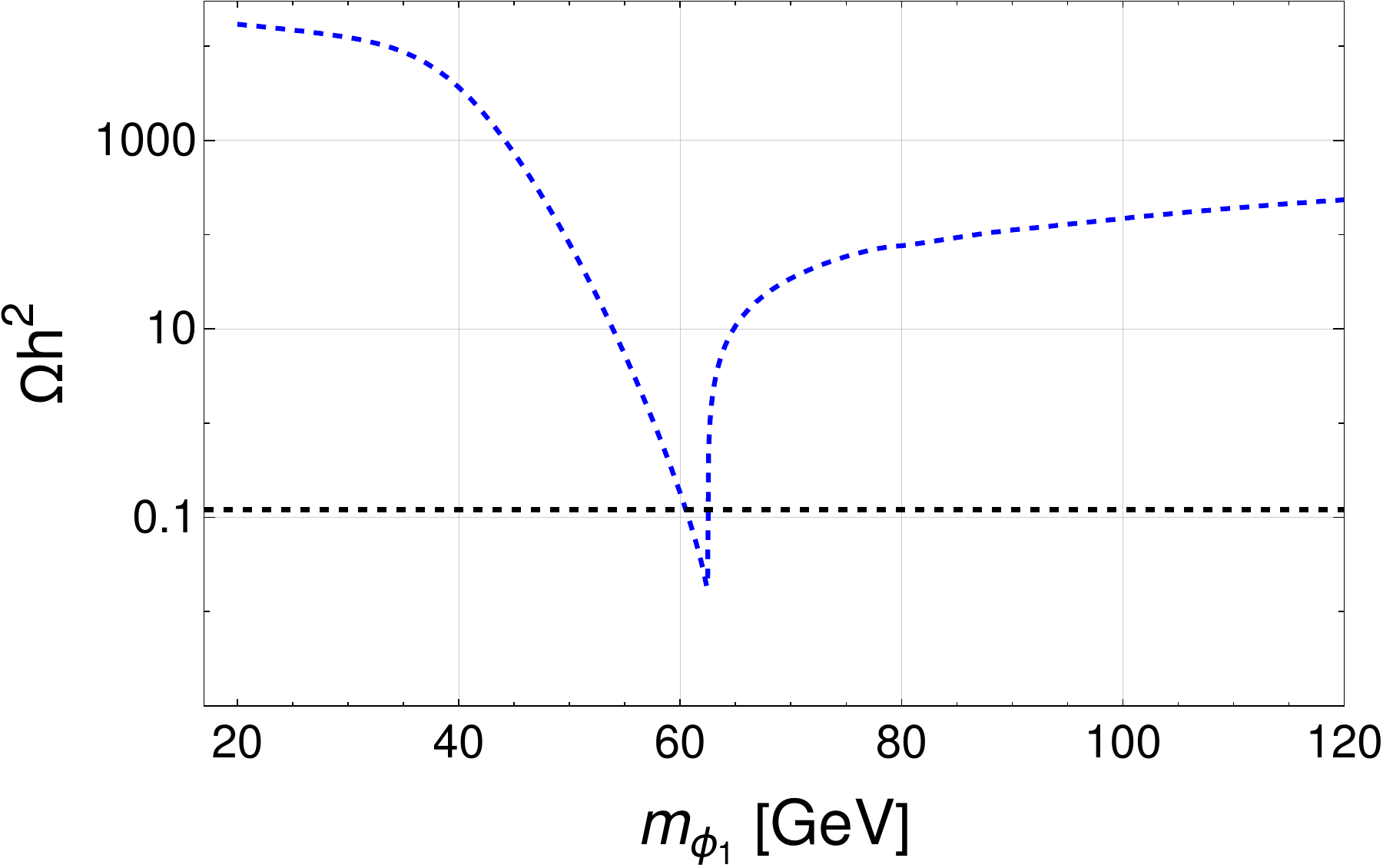}
    \includegraphics[width=0.48\textwidth]{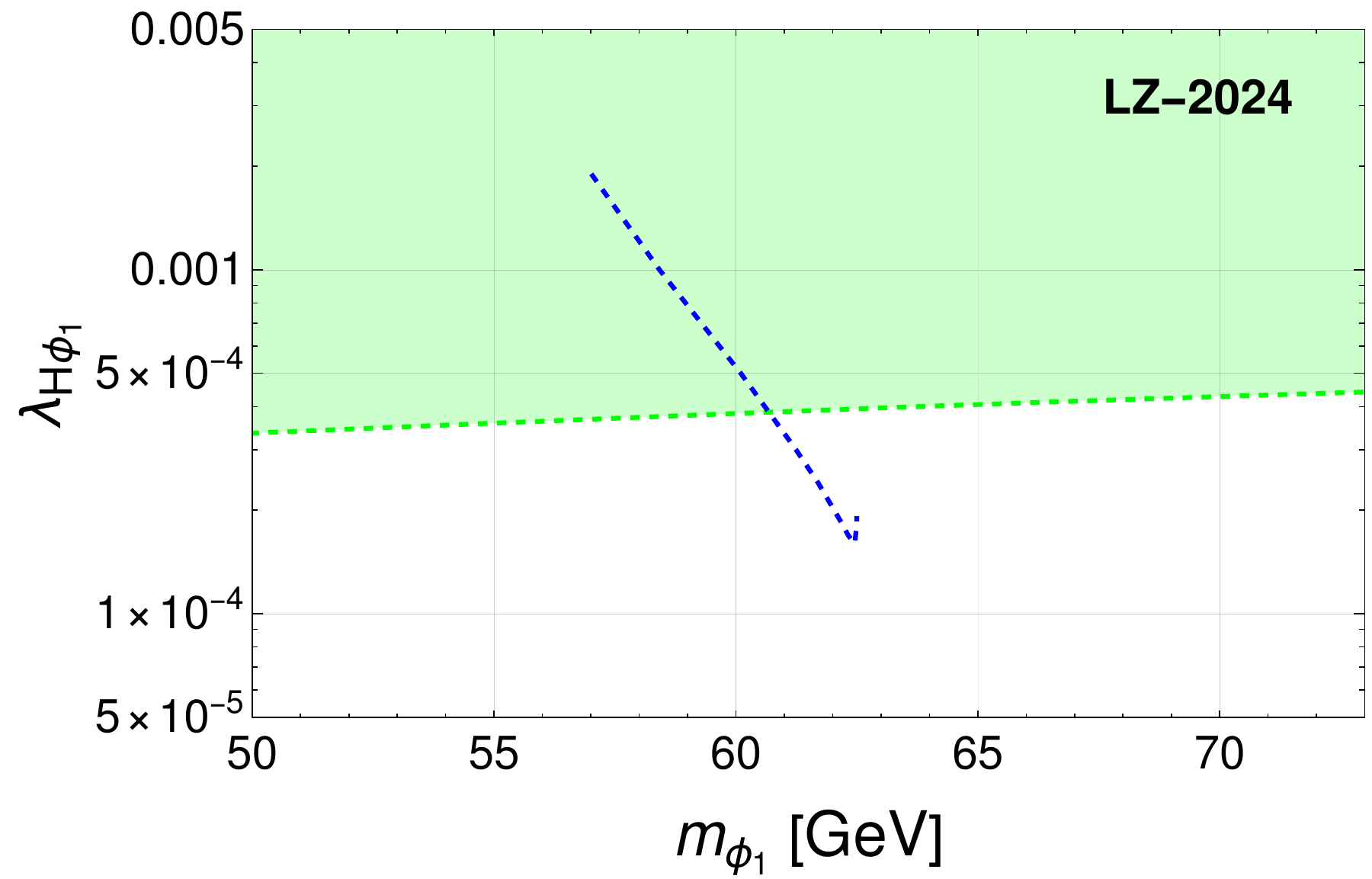}
    \caption{Left: Variation of DM relic abundance with dark matter mass \(m_{\phi_1}\). Right: Parameter space in \(\lambda_{H\phi_1}\) -- \(m_{\phi_1}\) plane for which the observed relic density is satisfied. The horizontal black dotted line in the left shows the current relic abundances $\Omega h^2 = 0.12$. The green shaded region in the right panel shows the bound from LZ-2024 \cite{LZ:2024zvo}.}
    \label{fig:relic-resonance}
\end{figure}

The left panel of Fig.\,\ref{fig:relic-resonance} shows the variation of the relic abundance as a function of the scalar mass \(m_{\phi_1}\). A sharp resonance appears near \(m_{\phi_1} \sim m_h/2\), where the annihilation cross-section is resonantly enhanced via the \(s\)-channel Higgs exchange. This allows the model to reproduce the correct relic abundance even for very small values of the Higgs portal coupling \(\lambda_{H\phi_1}\).

The right panel of Fig.\,\ref{fig:relic-resonance} displays the allowed region in the \((m_{\phi_1}, \lambda_{H\phi_1})\) parameter space consistent with the observed relic abundance. The viable points lie very close to the Higgs resonance and require extremely small portal couplings, typically \(\lambda_{H\phi_1} \lesssim 10^{-4}\), in order to evade current direct detection bounds. We overlay the latest LZ exclusion limit to illustrate that only a narrow strip in parameter space remains viable. Such small couplings significantly limit the testability of the minimal model in upcoming experiments, although future detectors may further constrain or probe this region.

The key outcome of this analysis is that only a narrow region near the Higgs resonance remains compatible with all current constraints, and it requires tiny portal couplings that make the model difficult to test through traditional means.

To go beyond this minimal scenario, we explore whether extensions of the scalar sector can have observable implications for early-Universe cosmology. As a first step, we introduce an additional real scalar singlet, also odd under a \(\mathbb{Z}_2\) symmetry, to study the impact on electroweak phase transition dynamics. We then consider a further extension by adding a third singlet scalar, with the goal of investigating the gravitational wave signals generated by a first-order electroweak phase transition—an essential ingredient for electroweak baryogenesis and a potential source of detectable stochastic gravitational waves.


\subsection*{Scenario 1: Two Scalar Singlets – One DM Component}

In this scenario, we extend the Standard Model by introducing two real scalar singlets, \(\phi_1\) and \(\phi_2\), along with an additional discrete symmetry \(\mathcal{G} = \mathbb{Z}_2^{(1)} \otimes \mathbb{Z}_2^{(2)}\). Under this symmetry, the fields \(\phi_1\) and \(\phi_2\) carry charges \((-1, 1)\) and \((1, -1)\), respectively, while all Standard Model fields remain even under \(\mathcal{G}\).
Among these, $\phi_1$ is the lightest and hence stable, making it a viable dark matter (DM) candidate. The heavier scalar $\phi_2$, while also $\mathbb{Z}_2$-odd and stable, contributes only subdominantly to the present-day relic abundance.

A notable feature of this setup is the strategic placement of scalar masses near the Higgs resonance region, i.e., $m_{\phi_i}  \approx 62.5$~GeV. This is a well-motivated choice, as illustrated in Fig.\,\ref{fig:relic-resonance}, as it enhances the DM annihilation cross-section through an $s$-channel Higgs exchange. At resonance, the annihilation cross-section is resonantly amplified due to the propagator structure of the intermediate Higgs:
\[
\langle \sigma v \rangle_{\phi_1 \phi_1 \to \text{SM}} \propto \frac{\lambda_{H\phi_1}^2}{(4 m_{\phi_1}^2 - m_h^2)^2 + m_h^2 \Gamma_h^2},
\]
allowing the correct relic abundance to be obtained even for very small values of the portal coupling $\lambda_{H\phi_1}$. This is particularly important to evade stringent bounds from spin-independent direct detection experiments, since the DM-nucleon cross-section scales as $\sigma_{\text{SI}} \propto \lambda_{H\phi_1}^2$.

We consider three benchmark points that successfully reproduce the observed dark matter relic abundance as measured by Planck as well as satisfy the direct detection bound from LZ-2024 data. The dominant annihilation channels $\phi_1$ are as follows: for BP-1, $\phi_1 \phi_1 \to b\bar{b}$ (39.8\%), $c\bar{c}$ (2.93\%), and $\tau^+ \tau^-$ (1.91\%); for BP-2, $\phi_1 \phi_1 \to b\bar{b}$ (55.8\%), $c\bar{c}$ (4.10\%), and $\tau^+ \tau^-$ (2.68\%); and for BP-3, $\phi_1 \phi_1 \to b\bar{b}$ (29.9\%), $c\bar{c}$ (2.20\%), and $\tau^+ \tau^-$ (1.43\%). These annihilation modes efficiently deplete $\phi_1$ in the early universe, leading to the correct relic density. The second scalar, $\phi_2$, though phenomenologically active and possessing sizable annihilation cross-sections, contributes negligibly to the relic abundance due to resonance-enhanced annihilation near $m_{\phi_2} \approx m_h/2$.




The effective spin-independent(SI) nucleon dark matter interaction can be written as~\cite{Bhattacharya:2016ysw}

\begin{eqnarray}
   \sigma^{\rm eff}_{\rm SI} = \dfrac{\Omega_1}{\Omega_{\rm tot}} \sigma^{(1)}_{\rm SI}+ \sum_{i=2}^n\dfrac{\Omega_i}{\Omega_{\rm tot}} \left(\dfrac{m_{\phi_1}}{m_{\phi_i}} \right) \sigma^{(i)}_{\rm SI}. 
\label{eq:DD}
\end{eqnarray}
While the spin-independent scattering cross-section $\sigma^{(2)}_{\rm SI}$ can become sizable due to a large Higgs portal coupling, the associated relic abundance $\Omega_2$ is strongly suppressed near the resonance region. In fact, this suppression is a direct consequence of placing the second scalar $\phi_2$ near the Higgs resonance ($m_{\phi_2} \approx m_h/2$), where annihilation is highly efficient, leading to a drastically reduced relic density. As a result, the effective spin-independent cross-section $\sigma^{\rm eff}_{\rm SI}$, which is weighted by the relic abundance, remains well below the sensitivity of current direct detection experiments, thereby avoiding exclusion.
The spin-independent scattering cross-sections for $\phi_1$ lie in the range of $\mathcal{O}(10^{-49})~\text{cm}^2$ across all three benchmarks, keeping them well below current direct detection limits. In contrast, the cross-sections for $\phi_2$ are significantly larger, of the order $\mathcal{O}(10^{-43})~\text{cm}^2$, due to its sizable Higgs portal coupling. However, since $\phi_2$ contributes negligibly to the total relic abundance, its large cross-section does not lead to exclusion and instead offers a promising target for future direct detection experiments.
 We have summarized the parameter values for all three chosen benchmarks in Tab.\,\ref{tab:two} and corresponding results in Tab.\,\ref{tab:resdm_sce2}.
 This two-component scalar dark matter framework naturally decouples relic abundance generation from direct detection constraints. By positioning $\phi_2$ at the resonance, the model accommodates a larger Higgs portal coupling without conflicting with dark matter observations. This setup not only stabilizes the relic density of the lighter component ($\phi_1$) through resonance effects but also renders the second scalar $\phi_2$ phenomenologically accessible. As such, the model provides a viable and testable extension to minimal Higgs portal scenarios, opening new avenues for probing scalar dark matter in future detection experiments.
\begin{table}[ht]
    \centering
    \begin{tabular}{|c|c|c|c|c|c|c|c|}
    \hline 
       BP & $m_{\phi_1} $ \small [GeV] & $m_{\phi_2}$ \small [GeV] &  $\lambda_{H\phi_1}$ & $\lambda_{H\phi_2}$& $\lambda_{\phi_1} $ & $\lambda_{\phi_2}$ & $\lambda_{\phi_1\phi_2}$ \\
       & & &  &  & & & \\
         \hline
        BP-1   & 60.52 & $m_h/2$ & 0.0004 & 0.40 & 0.01 & 0.19  & 0.001\\
         & & &&&&&\\
        \hline
        BP-2 & 62.13 & $m_h/2$ & 0.0002 & 0.30  & 0.01 & 0.11  &0.001 \\
         & & &&&&&\\
        \hline
        BP-3 & 61.43 & $m_h/2$  & 0.0003  & 0.25 & 0.01 & 0.09  & 0.001 \\
         & & &&&&&\\
        \hline
    \end{tabular}
    \caption{Benchmark points for two singlet scalar extensions. }
    \label{tab:two}
\end{table}

\begin{table}[htbp]
\centering
\renewcommand{\arraystretch}{1.3}

\begin{tabular}{|c|c|c|c|}
\hline
\textbf{Observable} & \textbf{BP-1} & \textbf{BP-2} & \textbf{BP-3} \\
\hline
\multicolumn{4}{|c|}{\textbf{Relic abundance}} \\
\hline
$\Omega_1 h^2$ & $1.20 \times 10^{-1}$ & $1.20 \times 10^{-1}$ & $1.20 \times 10^{-1}$ \\
$\Omega_2 h^2$ & $1.07 \times 10^{-7}$ & $2.56 \times 10^{-7}$ & $2.94 \times 10^{-7}$ \\
\hline
\multicolumn{4}{|c|}{\textbf{Spin-independent cross-section [$\text{cm}^2$]}} \\
\hline
$\sigma_{\rm SI}^{(1)}$ & $ 3.81 \times 10^{-49}$ & $ 9.04 \times 10^{-50}$ & $ 2.08 \times 10^{-49}$ \\
$\sigma_{\rm SI}^{(2)}$ & $ 3.58 \times 10^{-43}$   & $ 2.01 \times 10^{-43}$  & $ 1.40 \times 10^{-43}$ \\
$\sigma_{\rm SI}^{\rm eff}$ & $ 6.90\times 10^{-49}$   & $5.17 \times 10^{-49}$ & $5.44 \times 10^{-49}$ \\
\hline
\multicolumn{4}{|c|}{\textbf{Annihilation channels (fractional contributions)}} \\
\hline
$\phi_1\phi_1 \to b\bar{b}$ & $3.98 \times 10^{-1}$ & $5.58 \times 10^{-1}$ & $2.99 \times 10^{-1}$ \\
$\phi_2\phi_2 \to b\bar{b}$ & $4.93 \times 10^{-1}$ & $3.33 \times 10^{-1}$ & $5.93 \times 10^{-1}$ \\
$<\sigma v>^{\rm tot} [\rm cm^3/s]$  & $2.58\times 10^{-29}$ & $1.29\times 10^{-28}$ & $6.54\times 10^{-29}$ \\
\hline
\end{tabular}

\caption{Summary of benchmark points illustrating spin-independent DM-nucleon cross-sections, relic abundances, and dominant annihilation channels for $\phi_1$ and $\phi_2$.  We present the neutron spin-independent cross-section $\sigma_{\rm SI}^{}$, as it is the larger of the two and provides a more conservative comparison with experimental bounds.
}
\label{tab:resdm_sce2}
\end{table}
\subsection*{Scenario 2: Three Scalar Singlets – One DM and Two Inert States}

Building upon the previous scenario with two scalar singlets, we further extend the model by introducing a third real singlet scalar field, \(\phi_3\). The discrete symmetry is correspondingly enlarged to \(\mathcal{G} = \mathbb{Z}_2^{(1)} \otimes \mathbb{Z}_2^{(2)} \otimes \mathbb{Z}_2^{(3)}\), under which the scalar fields transform as \(\phi_1 \sim (-1, 1, 1)\), \(\phi_2 \sim (1, -1, 1)\), and \(\phi_3 \sim (1, 1, -1)\). All Standard Model fields remain even under \(\mathcal{G}\). This symmetry structure ensures that each scalar field is stabilized independently, preventing unwanted mixing and allowing controlled interactions via the Higgs portal.

 Among these, $\phi_1$ is the lightest and stable, making it the sole dark matter (DM) candidate, whereas $\phi_2$ and $\phi_3$ are heavier inert scalars that influence the early universe thermal history and potential phase transition dynamics but contribute negligibly to the present-day DM relic density.

Following the same guiding principle as in the two-scalar case, the mass of $\phi_1$ is chosen close to the Higgs resonance, $\approx 62.5$~GeV, to benefit from resonant enhancement of annihilation cross-sections via $s$-channel Higgs exchange. This allows small Higgs portal couplings $\lambda_{H\phi_1}$ to achieve the correct relic density while keeping spin-independent DM–nucleon cross-sections well below experimental bounds. The dominant annihilation channels of $\phi_1$ are into $b\bar{b}$, $c\bar{c}$, and $\tau^+\tau^-$, with approximate contributions of $85\%$, $6\%$, and $4\%$, respectively. The heavier states $\phi_2$ and $\phi_3$ contribute only at the $\sim 2\%$ level through $b\bar{b}$ final states, with their effect thermally suppressed due to their instability.
The $\phi_2$ and $\phi_3$ fields are at Higgs resonance$\approx m_h/2$, leading to highly efficient annihilation. As a result, they contribute negligibly to the relic density. However, this efficient annihilation allows their Higgs portal couplings to be sufficiently large to produce significant phenomenological and cosmological effects.
The benchmark points selected for this scenario are shown in Tab.\,\ref{tab:three}. The additional input parameters—$m_{\phi_i}$ and $\lambda_{H\phi_i}$ for $i=2,3$—are taken to match those used for $\phi_2$ in Tab.\,\ref{tab:two}, ensuring a consistent comparison with the two-scalar case.

\begin{table}[ht]
    \centering
    \begin{tabular}{|c|c|c|c|c|c|c|}
    \hline 
       BP & $m_{\phi_1} $ \small [GeV] & $\lambda_{H\phi_1}$ &$\lambda_{\phi_1}$ & $\lambda_{\phi_i}$  & $\lambda_{\phi_1\phi_i}$   &$\lambda_{\phi_2\phi_3}$  \\
       &  &  & & ($i=2,3$) & ($i=2,3$)   &  \\
         \hline
        BP-1   & 61.70  & 0.0002  & 0.01 & 0.29 &  0.001  & 0.16    \\
         & & &&&&\\
        \hline
        BP-2 & 61.14  & 0.0003 & 0.01 & 0.16 & 0.001    & 0.10  \\
         & & &&&&\\
        \hline
        BP-3 & 62.06  & 0.0002 & 0.01 &  0.10 & 0.001    & 0.11   \\
         & & &&&&\\
        \hline
    \end{tabular}
    \caption{Benchmark points for three singlet scalar extensions. For this scenario, the other input values $m_{\phi_i}$ and $\lambda_{H\phi_i}$ for $i=2,3$ are same $m_{\phi_2}$, $\lambda_{H\phi_2}$ respectively, given in Tab.\,\ref{tab:two}. }
    \label{tab:three}
\end{table}

In BP-1, the DM candidate mass is $m_{\phi_1} = 60.4$~GeV, which lies very close to the Higgs resonance. A very small portal coupling $\lambda_{H\phi_1} = 0.0004$ is sufficient to deplete the relic abundance to the observed level, with the advantage of keeping the spin-independent cross-section well below the exclusion limits from XENON1T and LUX-ZEPLIN (LZ). The quartic coupling for the inert scalars, $\lambda_{\phi_i} = 0.29$, is relatively large and may affect the scalar thermal potential and dynamics of the electroweak phase transition. However, the interscalar mixing terms $\lambda_{\phi_1\phi_i} = 0.001$ remain small, ensuring minimal backreaction on $\phi_1$ evolution. We consider three benchmark points that successfully reproduce the observed dark matter relic abundance as measured by Planck, while remaining consistent with the direct detection limits from LZ-2024. The dominant annihilation channels for the main dark matter component $\phi_1$ are: for BP-1, $\phi_1 \phi_1 \to b\bar{b}$ (25.9\%), $c\bar{c}$ (1.90\%), and $\tau^+ \tau^-$ (1.24\%); for BP-2, $\phi_1 \phi_1 \to b\bar{b}$ (17.0\%), $c\bar{c}$ (1.25\%), and $\tau^+ \tau^-$ (0.82\%); and for BP-3, $\phi_1 \phi_1 \to b\bar{b}$ (27.9\%), $c\bar{c}$ (2.06\%), and $\tau^+ \tau^-$ (1.34\%). These channels are sufficient to deplete $\phi_1$ in the early universe, yielding the correct relic abundance. The scalars $\phi_2$ and $\phi_3$, denoted collectively as $\phi_i$, also undergo annihilations into the same final states with sizable branching ratios—$b\bar{b}$ (30–36\%), $c\bar{c}$ (2–3\%), and $\tau^+\tau^-$ (1.5–1.7\%)—but their relic contributions remain negligible due to efficient annihilation near the Higgs resonance, $m_{\phi_i} \approx m_h/2$.
 The relic contribution of $\phi_1$ remains dominant ($\Omega_1 h^2 \sim 0.12$), while $\phi_2$ and $\phi_3$ each contribute at the $\mathcal{O}(10^{-7})$ level. The annihilation channels for both $\phi_1$ and $\phi_i$ are again dominated by $b\bar{b}$, $c\bar{c}$, and $\tau^+\tau^-$ with similar branching patterns. The spin-independent DM–nucleon cross-sections for $\phi_1$ range from $10^{-49}-10^{-48}$~$\text{cm}^2$ across all benchmarks and remain comfortably below current experimental bounds. 

The three-scalar setup thus generalizes the two-scalar model by introducing further flexibility in scalar interactions and thermal history, while maintaining compatibility with dark matter and collider constraints. It opens the door for richer dynamics in early-universe cosmology and electroweak phase transition studies without disrupting the observed dark matter phenomenology.

\begin{table}[htbp]
\centering
\renewcommand{\arraystretch}{1.3}
\begin{tabular}{|c|c|c|c|}
\hline
\textbf{Observable} & \textbf{BP-1} & \textbf{BP-2} & \textbf{BP-3} \\
\hline
\multicolumn{4}{|c|}{\textbf{Relic abundance}} \\
\hline
$\Omega_1 h^2$ & $1.19 \times 10^{-1}$ & $1.19 \times 10^{-1}$ & $1.21 \times 10^{-1}$ \\
$\Omega_i h^2$ & $1.27 \times 10^{-7}$ & $1.99 \times 10^{-7}$ & $3.52 \times 10^{-7}$ \\
\hline
\multicolumn{4}{|c|}{\textbf{Spin-independent cross-section [$\text{cm}^2$]}} \\
\hline
$\sigma_{\rm SI}^{(1)}$ & $9.17 \times 10^{-50}$ & $2.10 \times 10^{-49}$ & $9.06 \times 10^{-50}$ \\
$\sigma_{\rm SI}^{(i)}$ & $3.58 \times 10^{-43}$   & $2.01 \times 10^{-43}$  & $1.40 \times 10^{-43}$ \\
$\sigma_{\rm SI}^{\rm eff}$ & $8.45 \times 10^{-49}$   & $ 8.68 \times 10^{-49}$  & $ 8.97 \times 10^{-49}$ \\
\hline
\multicolumn{4}{|c|}{\textbf{Annihilation channels (fractional contributions)}} \\
\hline
$\phi_1\phi_1 \to b\bar{b}$ & $2.59 \times 10^{-1}$ & $1.70 \times 10^{-1}$ & $2.79 \times 10^{-1}$ \\
$\phi_i\phi_i \to b\bar{b}$ & $3.16 \times 10^{-1}$ & $3.61 \times 10^{-1}$ & $3.06 \times 10^{-1}$ \\
$<\sigma v>^{\rm tot} [\rm cm^3/s]$  & $5.98\times 10^{-29}$ & $7.14\times 10^{-29}$ & $1.82\times 10^{-28}$ \\
\hline
\end{tabular}
\caption{Benchmark points in the three-scalar model showing spin-independent DM-nucleon cross-sections, relic abundances, and dominant annihilation channels. Here, $\phi_i$ collectively denotes the subdominant scalars $\phi_2$ and $\phi_3$, while $\phi_1$ is the dominant dark matter component. We present the neutron spin-independent cross-section $\sigma_{\rm SI}$ throughout, as it is the larger of the two and provides a more conservative comparison with experimental bounds. }
\label{tab:3scalar_benchmarksres}
\end{table}
The spin-independent scattering cross-sections for $\phi_1$ remain in the range of $\mathcal{O}(10^{-49})~\text{cm}^2$ across all benchmarks, safely below the LZ-2024 exclusion limits, while the heavier scalars $\phi_2$ and $\phi_3$ exhibit larger cross-sections of $\mathcal{O}(10^{-43})~\text{cm}^2$ due to their sizable Higgs portal couplings. However, since their relic contributions are strongly suppressed by resonance-enhanced annihilation, the effective cross-section remains below current detection thresholds, allowing all benchmark points to evade exclusion.

A comprehensive summary of the detailed numerical results for the scenario are provided in Tab.\,\ref{tab:3scalar_benchmarksres}.


\begin{figure}[t]
    \centering
\includegraphics[width=0.495\textwidth]{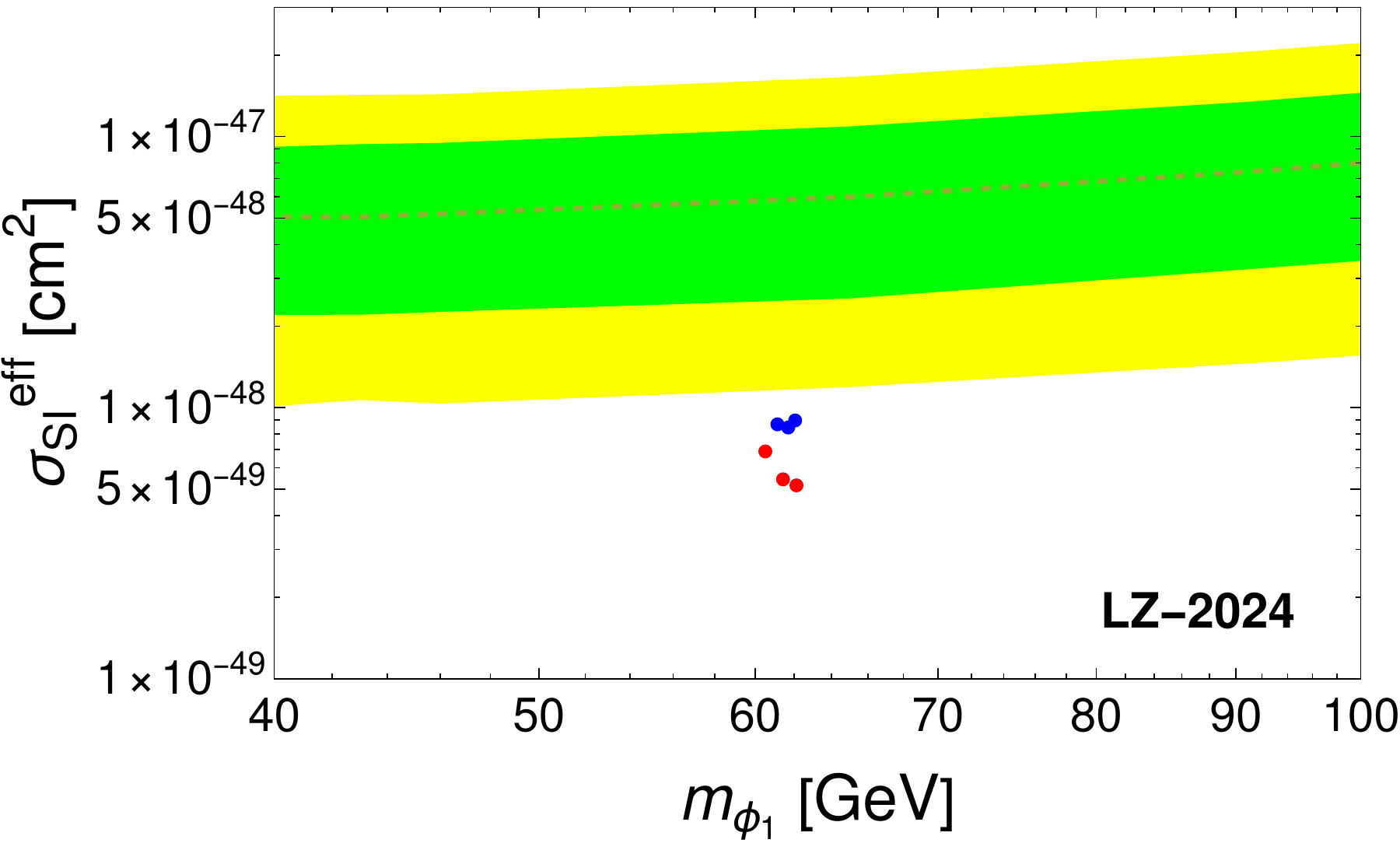}
 \includegraphics[width=0.47\textwidth]{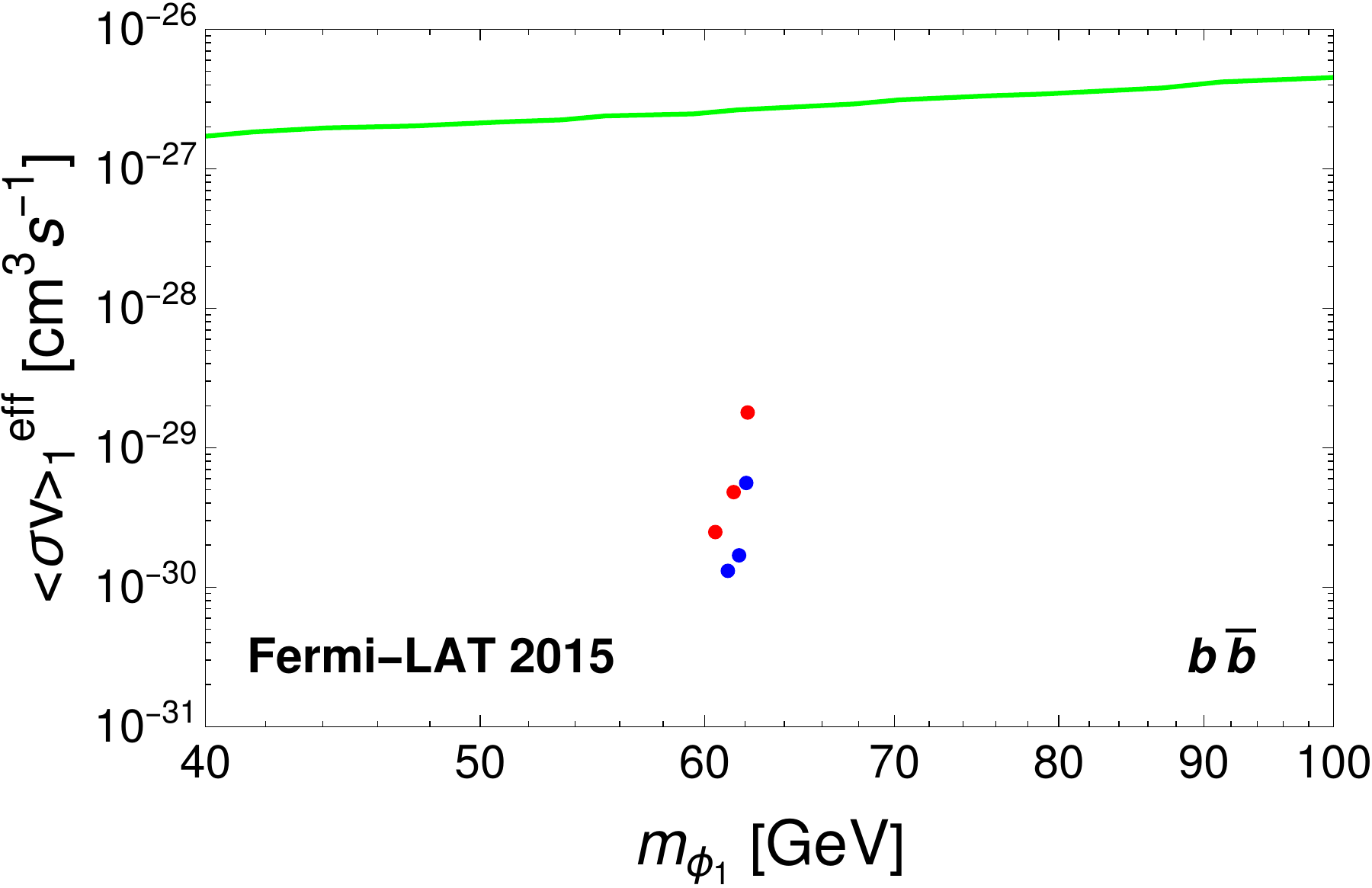}
 \\
 \hspace{30pt}(a)  \hspace{180pt}(b)
    \caption{(a): Variation of nucleon dark matter effective spin-independent direct detection cross section with dark matter mass. Red points correspond to the three benchmark points in the two-scalar scenario, while blue points represent the three benchmarks for the three-scalar scenario. The shaded green and yellow bands indicate the $1\sigma$ and $2\sigma$ sensitivity regions of the LUX-ZEPLIN (LZ) 2024 experiment, respectively. The experimental exclusion band is adapted from Ref.~\cite{LZ:2024zvo}. 
    (b): Thermally averaged annihilation cross-section \(\langle \sigma v \rangle\) of the dark matter candidate \(\phi_1\) into \(b\bar{b}\) final states as a function of the DM mass \(m_{\phi_1}\). The red (blue) points correspond to benchmark scenarios from the two-scalar (three-scalar) models. The solid green curve represents the 95\% C.L. upper limit on \(\langle \sigma v \rangle\) from Fermi-LAT gamma-ray observations of dwarf spheroidal galaxies~\cite{Fermi-LAT:2015att}.
    }
    \label{fig:LZsigmaEff}
\end{figure}

We note that in the two-singlet scalar scenario, the dark matter relic abundance is independent of the quartic couplings $\lambda_{\phi_i}$ for $i = 1,2$, while in the three-singlet extension it remains independent of $\lambda_{\phi_i}$ ($i = 1,2,3$) and $\lambda_{\phi_2 \phi_3}$. However, the relic abundance exhibits a mild dependence on the couplings $\lambda_{\phi_1 \phi_i}$, with $i = 2$ for the two-scalar case and $i = 2,3$ for the three-scalar scenario. 

In contrast, the spin-independent dark matter–nucleon direct detection cross section is solely governed by the Higgs portal couplings $\lambda_{H\phi_i}$ and the corresponding scalar masses $m_{\phi_i}$. Consequently, the commonly presented exclusion plots in the $\sigma_{\rm SI}$ vs.~$m_{\rm DM}$ plane can equivalently be recast in the $\lambda_{H\phi_1}$ vs.~$m_{\phi_1}$ plane, offering a more direct view of the underlying parameter space. In Fig.\,\ref{fig:LZsigmaEff} (a), we use the spin-independent cross-section for the neutron only—specifically the larger of the two—as it yields a more conservative estimate.

As illustrated in Fig.\,\ref{fig:LZsigmaEff}(a), all six benchmark points—three for each scenario—yield effective cross sections well below the $2\sigma$ exclusion limit of the latest LZ-2024 data. Notably, these benchmark configurations lie close to the projected sensitivity frontier, indicating that they may become testable in the upcoming generation of direct detection experiments.

As shown in Fig.\,\ref{fig:LZsigmaEff}(a), the benchmark points for the three-scalar scenario (indicated in blue) lie closer to the LZ-2024 direct detection exclusion region compared to those from the two-scalar case (in red). This behavior arises because the effective spin-independent cross-section, defined in Eq.\,\ref{eq:DD}, which accumulates contributions from all DM scalar components~\cite{Bhattacharya:2016ysw}. 
This becomes particularly relevant in scenarios where multiple dark sector particles have comparable masses. In such cases, the detector's nuclear recoil response is largely insensitive to individual masses, making it difficult to distinguish between components. Thus, for practical purposes, their scattering effects must be added. In the three-scalar setup, each inert scalar $\phi_i$ contributes only a small fraction to the relic abundance because of resonance-enhanced annihilation near $m_{\phi_i} \approx m_h/2$; however, their combined contribution to the effective spin-independent cross-section $\sigma_{\rm SI}^{\rm eff}$ remains significant.

This is particularly important because, at the resonance, the product $\Omega_i h^2 \times \sigma_{\rm SI}^{(i)}$ remains approximately constant over a broad range of Higgs portal couplings. As a result, adding more scalars with $\mathcal{O}(1)$ Higgs portal couplings—does not reduce the effective cross-section proportionally. This constrains the number of singlet scalars that can be introduced near the resonance without violating direct detection limits. Therefore, while multiple scalars may help strengthen the electroweak phase transition, such extensions must be carefully balanced to remain consistent with current experimental constraints.

Next, we analyze the indirect detection prospects of dark matter in this multicomponent setup. The effective thermally averaged annihilation cross-section relevant for indirect signals from a given component $\phi_i$ is determined by its fractional contribution to the total relic density. It is given by~\cite{Ghosh:2022rta, Bhattacharya:2019fgs}:
\begin{equation}
\langle \sigma v \rangle^{\rm eff}_i = \left( \frac{\Omega_i}{\Omega_{\rm tot}} \right)^2 \left(\dfrac{m_{\phi_i}}{m_{\phi_i}+m_{\phi_j}} \right)^2 \langle \sigma v \rangle_i\,,~~(j\neq i)
\end{equation}
where $\langle \sigma v \rangle_i$ denotes the intrinsic thermally averaged annihilation cross-section for $\phi_i$, and $\Omega_i / \Omega_{\rm tot}$ represents its relative relic abundance. This relation clearly indicates that components with subdominant relic densities (e.g., $\phi_j$ with $j \ne 1$) exhibit a strong suppression in their effective annihilation cross-sections, even if their individual rates are sizable. Consequently, their contribution to indirect detection signals is significantly reduced.

Fig.\,\ref{fig:LZsigmaEff}(b), shows the thermally averaged annihilation cross-section \(\langle \sigma v \rangle\) for the dark matter candidate \(\phi_1\) into \(b\bar{b}\) final states, overlaid with the Fermi-LAT 2015 upper bound derived from observations of dwarf spheroidal galaxies~\cite{Fermi-LAT:2015att}. The benchmark points for both the two-scalar (red) and three-scalar (blue) scenarios lie well below the exclusion limit, indicating consistency with current indirect detection constraints. Among all possible final states, annihilation into \(b\bar{b}\) dominates due to the large Yukawa coupling of the bottom quark. Other channels such as \(c\bar{c}\), \(\tau^+ \tau^-\), and lighter fermions have significantly smaller branching fractions and thus contribute only subdominantly to the total annihilation rate. Consequently, their impact on the predicted gamma-ray flux is negligible, and the \(b\bar{b}\) final state provides the most stringent constraint on the model from Fermi-LAT data.



To ensure compatibility with collider bounds—especially constraints on the invisible decay width of the Higgs boson—we examine the collider implications of our scenarios in the following section.


\section{Collider Constraints}
\label{sec:collider}
Collider searches offer powerful probes of Higgs-portal dark matter scenarios. We focus on two key constraints within our multi-scalar singlet framework: the Higgs invisible decay width and monojet plus missing transverse momentum (MET) searches at the LHC. In the following subsections, we present our analysis and interpret existing experimental results to evaluate their impact on the model's parameter space.

\subsection{Higgs Invisible Decay}
\begin{figure}[tb]
    \centering
    \includegraphics[width=0.46\linewidth]{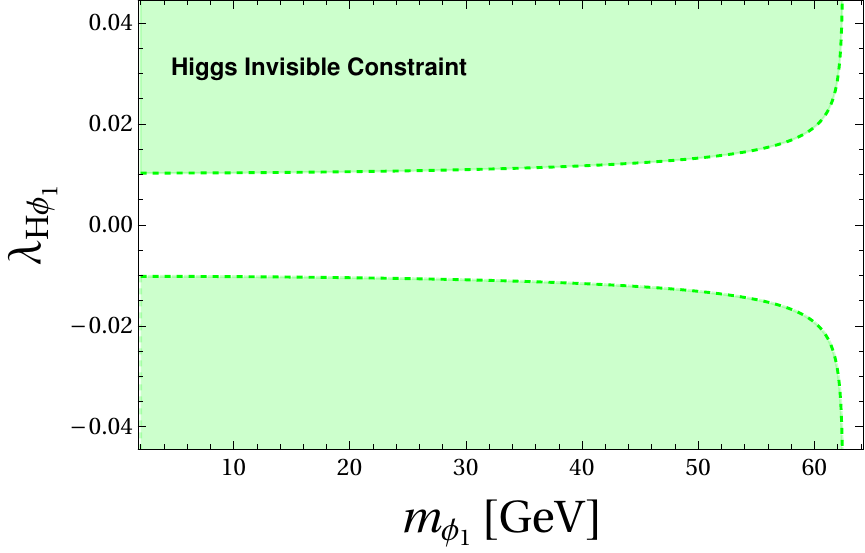} \hspace{2mm}
    \includegraphics[width=0.46\linewidth]{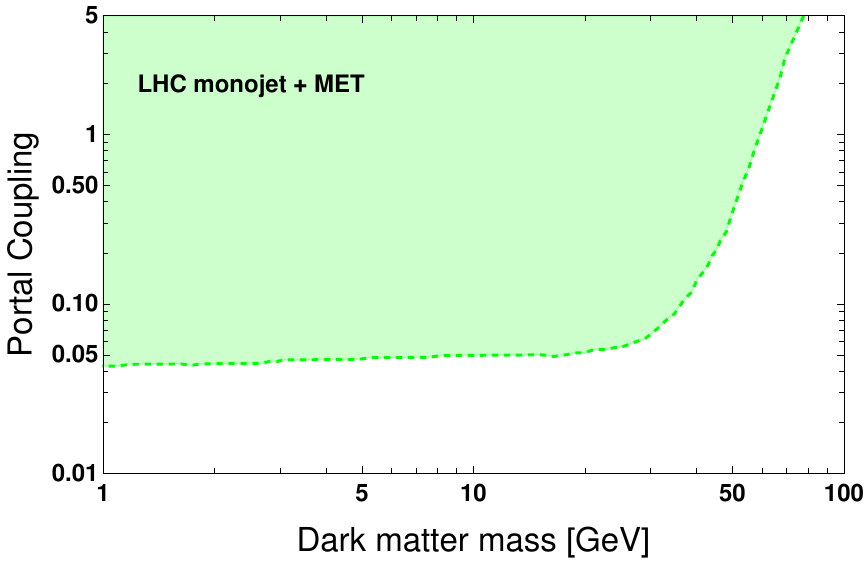}
    \caption{Left panel: Constraints on the Higgs portal coupling as a function of dark matter mass, derived from limits on the Higgs invisible decay branching ratio. Right panel: Exclusion limits at $95\%$ confidence level from LHC monojet searches for scalar dark matter.}
    \label{fig:HiggsLHC}
\end{figure}

In our model, the Higgs boson production rate at the LHC remains identical to that in the SM. However, for dark matter masses below half the Higgs mass, the experimental bound on the Higgs invisible branching ratio~\cite{ATLAS:2023tkt} imposes constraints on the dark matter mass versus Higgs portal coupling plane. For $m_{\phi_i}< \frac{m_h}{2}$, the partial decay width of the Higgs boson into a single pair of scalar particles is given below~\cite{Kanemura:2010sh}
\begin{equation}
\Gamma(h \to \phi_i\phi_i) = \frac{\lambda_{H\phi_i}^2 v^2}{32\,\pi\, m_h} \left(1 - \frac{4\, m_{\phi_i}^2} {m_h^2}\right)^{1/2}~.
\end{equation}
Although the model contains multiple BSM scalars, for the benchmark points listed in Tab.\,\ref{tab:two} and \ref{tab:three}, only the Higgs decay into a pair of $\phi_1$ is kinematically allowed. The corresponding contribution to the invisible decay width is
\begin{equation}
\Gamma_{\text{inv}}^{\text{BSM}} =\Gamma(h \to \phi_1\phi_1)~.
\end{equation}
The branching ratio for Higgs invisible decays is
\begin{equation}
\text{BR}(h \to \text{inv}) = \frac{\Gamma_{\text{inv}}^{\text{BSM}}}{\Gamma_h^{\text{SM}} + \Gamma_{\text{inv}}^{\text{BSM}}} = \frac{\Gamma(h \to \phi_1\phi_1)}{\Gamma_h^{\text{SM}} + \Gamma(h \to \phi_1\phi_1)}~,
\end{equation}
where $\Gamma_h^{\text{SM}} \approx 4.07\,\text{MeV}$ is the SM Higgs width. Substituting explicitly for $\Gamma(h \to \phi_1\phi_1)$, we find
\begin{equation}
\text{BR}(h \to \text{inv}) = \frac{\lambda_{H\phi_1}^2 v^2}{ \lambda_{H\phi_1}^2 v^2 + 32\pi m_h \Gamma_h^{\text{SM}} \left(1 - \frac{4m_{\phi_1}^2}{m_h^2}\right)^{-1/2}}~~.
\end{equation}
Assuming a branching ratio of $\text{BR}(h \to \text{inv}) = 11\%$~\cite{ATLAS:2023tkt}, the constraints on the Higgs portal coupling as a function of the dark matter mass are presented in the left panel of Fig.\,\ref{fig:HiggsLHC}. The filled green regions indicate the parameter space excluded by the bound on the Higgs invisible decay width. Consequently, all the representative benchmark points considered in this study remain consistent with current Higgs invisible decay constraints.

\subsection{Monojet searches at the LHC}

\begin{table}[tb!]
	\begin{center}
		\small
			\begin{tabular}{|c|c|c|c|c|}
				\hline
			 &   \multicolumn{4}{|c|}{$\sigma(pp \to \phi_1 \phi_1 j) + \sigma(pp \to \phi_2 \phi_2 j)$}  \\
				\cline{2-5}
				  & $p_T^j \geq 30$ GeV & $p_T^j \geq 50$ GeV &  $p_T^j \geq 100$ GeV & $p_T^j \geq 250$ GeV\\ 
				\hline
                 \multirow{2}{*}{BP1 ($ |\eta_j| <5.0 $)}  & \multirow{2}{*}{$5.38^{+11.9\%}_{-10.3\%}$ pb} & \multirow{2}{*}{$2.93^{+12.9\%}_{-11.0\%}$ pb} &  \multirow{2}{*}{$0.943^{+14.7\%}_{-12.2\%}$ pb} & \multirow{2}{*}{$0.098^{+17.9\%}_{-14.3\%}$ pb}\\ 
                & & & & \\
                   \hline
                    \multirow{2}{*}{BP1 ($ |\eta_j| <2.4 $)}  & \multirow{2}{*}{$4.15^{+11.5\%}_{-10.1\%}$ pb} & \multirow{2}{*}{$2.37^{+12.6\%}_{-10.8\%}$ pb} &  \multirow{2}{*}{$0.817^{+14.6\%}_{-12.1\%}$ pb} & \multirow{2}{*}{$0.094^{+17.8\%}_{-14.3\%}$ pb}\\
                & & & & \\
                    \hline
\multirow{2}{*}{BP2 ($|\eta_j| < 2.4$)} & \multirow{2}{*}{$2.32^{+11.6\%}_{-10.2\%}$ pb} & \multirow{2}{*}{$1.34^{+12.6\%}_{-10.8\%}$ pb} & \multirow{2}{*}{$0.461^{+14.5\%}_{-12.1\%}$ pb} & \multirow{2}{*}{$0.053^{+17.8\%}_{-14.3\%}$ pb} \\
 & & & & \\
                    \hline
			\end{tabular} 
		\caption{The cross sections for dark matter pair production in association with a jet at the 13~TeV LHC are presented for two representative benchmark points, BP1 and BP2, from Tab.~\ref{tab:two}, with varying cuts on the jet transverse momentum \(p_T^j\). The first two rows correspond to BP1 with jet pseudorapidity ranges \( |\eta_j| < 5.0 \) and \( |\eta_j| < 2.4 \), respectively, while the third row corresponds to BP2 with \( |\eta_j| < 2.4 \). The superscripts and subscripts denote the renormalization and factorization scale uncertainties (in percentage) associated with the total cross section.}
		\label{tab:crosssection}
	\end{center}
\end{table}

All the BSM particles in our model are singlets under the SM gauge group and serve as dark matter candidates with different masses. Although the BSM scalars have varying masses, they cannot decay into the lightest scalar (denoted as $\phi_1$) due to the discrete $\mathcal{G}$ symmetry imposed in the Lagrangian. As a result, cascade decays of heavier BSM particles into the lightest dark matter and visible SM particles are forbidden. Consequently, monojet plus MET searches provide a promising probe of this simplified model at colliders. In contrast, more complex dark matter scenarios involving heavier dark sector particles and cascade decays can lead to more varied collider signatures, such as boosted three-pronged top fat jets, two-pronged fat jets, or fat jets accompanied by energetic charged leptons and significant MET~\cite{Ghosh:2021noq,Ghosh:2022rta,Ghosh:2023xhs,Ghosh:2024boo,Ghosh:2024nkj,Ghosh:2025agw}.

In proton-proton collisions, dark matter particles can be pair-produced via Higgs-mediated processes, often accompanied by initial-state radiation (ISR) of a high-$p_T$ jet, 
 \begin{equation}
    p p \to h^* + j \to \phi_i \phi_i + j~.
 \end{equation}
The pair production of dark matter particles can proceed via an on-shell Higgs boson if $m_{\phi_i}\leq m_h/2$. However, for $m_{\phi_i} > m_h/2$, the process can still contribute to the monojet plus MET final state through an off-shell Higgs boson. Therefore, in our model $\phi_i \phi_i + j,~~(i=1,2,..)$ final states must be included in the signal prediction. Terms like $\phi_1 \phi_2 + j$  is forbidden, since the interaction vertex $h\phi_i\phi_k~(i\neq k)$ is forbidden.

The model Lagrangian and the effective Higgs--gluon--gluon interaction ($\mathcal{L}_{\text{eff}}$) are implemented in \textsc{FeynRules}~\cite{Christensen:2008py,Alloul:2013bka} to generate the Universal FeynRules Output (\textsc{UFO})~\cite{Degrande:2011ua} model file. This \textsc{UFO} file is then used in the \textsc{MadGraph5\_aMC@NLO} framework~\cite{Alwall:2014hca,Frederix:2018nkq} to compute the cross sections and their associated scale uncertainties. The effective Lagrangian describing the coupling of the Higgs boson to gluons via a top-quark loop can be written as~\cite{Spira:1995rr}:
\begin{equation}
\mathcal{L}_{\text{eff}} = -\frac{1}{4} G_H \, G^{a\,\mu\nu} G^a_{\mu\nu} \, h,
\end{equation}
where \( G^a_{\mu\nu} \) is the gluon field strength tensor, \( h \) is the physical Higgs field, and the effective coupling \( G_H \) is given by~\cite{FeynRules_model}:
\begin{equation}
G_H = \frac{\alpha_s}{3\pi v} \left[1 + \frac{7}{30}x + \frac{2}{21}x^2 + \frac{26}{525}x^3 \right],
\end{equation}
with \( x = \left( \frac{m_h}{2 m_t} \right)^2 \), \( v \approx 246\,\text{GeV} \) being the Higgs vacuum expectation value, and \( m_t \) the top quark mass.
The cross sections for dark matter pair production with an associated jet at the 13~TeV LHC are presented in Tab.\,\ref{tab:crosssection} for two representative benchmark points, with varying jet transverse momentum cuts, using the \textsc{NNPDF3.0} parton distribution function (\textsc{PDF}) set~\cite{NNPDF:2014otw}. The cross sections are quoted for the central values of the renormalization and factorization scales, both set to the partonic centre-of-mass energy ($\sqrt{\hat{s}}$). To estimate theoretical uncertainties, the renormalization and factorization scales are varied independently by factors of $1/2$, $1$, and $2$ with respect to the central value, resulting in nine distinct scale combinations. The superscripts and subscripts in the table indicate the uncertainty envelope derived from these scale variations. 

As the dark matter particles escape the detector without interacting, these events are characterized by a energetic jet recoiling against a large MET. By analyzing the kinematic distributions of such events, LHC experiments placed bounds on the Higgs portal coupling strength as a function of the dark matter mass, considering only one dark matter candidate. The study~\cite{Ellis:2017ndg} includes monojet searches of CMS~\cite{CMS:2014jvv} and ATLAS~\cite{ATLAS:2014hqe,ATLAS:2015qlt,ATLAS:2016bek} at both $\sqrt{s}=8$ TeV and 13 TeV, analyzed using the \textsc{CheckMATE$_2$}~\cite{Dercks:2016npn,deFavereau:2013fsa,Cacciari:2011ma,Cacciari:2005hq} framework, and derives $95\%$ confidence level exclusion contour in the mass–coupling plane for the Higgs portal model. The corresponding contour of their analysis, considering Higgs decays into a pair of scalar dark matter candidates, is shown in the right panel of Fig.\,\ref{fig:HiggsLHC}. The filled green regions represent the corresponding excluded regions.
From this panel, for a dark matter mass of $m_h/2$, current LHC monojet plus MET searches impose an upper limit on the Higgs portal coupling of  $\lesssim 1.4$. Consequently, all the benchmark points considered in this study remain allowed under existing LHC constraints.

\section{Electroweak Phase Transition} 
\label{sec:ewpt}
To explore the electroweak phase transition (EWPT) in the early universe, we begin by studying the effective potential at finite temperature. We construct the analysis by first writing down the one-loop scalar potential at zero temperature and then adding thermal corrections, which play a critical role in determining the phase transition’s characteristics. The subsequent section is devoted to the study of stochastic GW production, with an emphasis on their prospects for detection in future experiments.
\subsection{One-loop effective potential}


The one-loop correction to the effective potential at zero temperature, known as the Coleman-Weinberg (CW) potential
using the $\overline{MS}$ scheme, this potential takes a specific form as given in~\cite{Coleman:1973jx,Quiros:1999jp}.
\begin{eqnarray}\label{eq:CWpot}
    V_{\rm 1-loop}^{\rm CW} (h,\phi_1,..\phi_n)
= \frac{1}{64\pi^2} \sum_{i}n_i m_i^4(h,\phi_i) \left[ \log \left( \frac{m_i^2(h,\phi_i)}{\mu^2} \right) - C_i \right],
\end{eqnarray}
where $m_i$'s are the field dependent masses of $i$th field, $\mu$ is the renormalization scale of the theory, and $C_i= 3/2$ for scalars and fermions and $C_i= 5/6$ for gauge bosons \cite{Wainwright:2011kj}. The degrees of freedom for the fermions and bosons are given by:
\begin{eqnarray}\label{eq:degrees}
n_W = 6,\,\,\ n_Z = 3,\,\,\ n_{h} = 1,\,\,\, n_{\phi_i} = 1,\,\  n_{\chi_{1,2,3}}=1,\,\,\ n_q = -12,\,\,\, n_\ell = -4.
\end{eqnarray}
%
Here, $q$ and $\ell$ denote the quarks and charged leptons, respectively. 

The mass-squared terms entering the CW potential are functions of the background fields. The field-dependent mass-squared matrix elements for the CP-even scalar sector can be written as:
\begin{align}\label{eq:masses}
&m_{hh}^2 (h,\phi_1,..\phi_n)\,\, \equiv \frac{\partial^2 V_0}{\partial h^2}=\mu_H^2 +3 \lambda_H h^2 + \sum_i^n\frac{\lambda_{H\phi_i} \phi_i^2}{2} \\
&m_{\phi_i\phi_i}^2(h,\phi_1,..\phi_n) \equiv \frac{\partial^2 V_0}{\partial \phi_i^2}=\mu_{\phi_i}^2 + 3 \lambda_{\phi_i} \phi_i^2+\frac{\lambda_{H\phi_i} h^2}{2}  + \sum_{j>i}^n \frac{\lambda_{ij}}{2} \phi_j^2, \\
&m_{h\phi_i}^2 (h,\phi_i) \,\, \equiv \frac{\partial^2 V_0}{\partial h \partial \phi_i}= \lambda_{H\phi_i} h \phi_i ,\quad m_{\phi_i\phi_j}^2 (\phi_i,\phi_j) \,\, \equiv \frac{\partial^2 V_0}{\partial \phi_i \partial \phi_j}=  \lambda_{ij} \phi_i \phi_j\label{eq:masses1} .
\end{align}
The fields-dependent masses of the CP-odd Goldstone bosons $\chi_i$ are
\begin{eqnarray}
    m_{\chi_i}^2(h,\phi_1,..\phi_n) =  \mu_H^2+\lambda_H h^2  + \sum_i^n \frac{\lambda_{H\phi_i} \phi_i^2}{2}.
\end{eqnarray}
The field-dependent masses of the fermions and gauge bosons are
\begin{eqnarray}
   m_i^2(h) = \frac{1}{2} y_i^2 h^2,\quad m_W^2(h) = \frac{1}{4} g^2 h^2,\quad m_Z^2(h) = \frac{1}{4} (g^2 + g^{\prime 2}) h^2
\end{eqnarray}
Here, $y_i=\sqrt{2}m_i/v$ is the Yukawa coupling of the corresponding fermion, and $g$ and $g^\prime$ are the gauge couplings corresponding to $SU(2)_L$ and $U(1)_Y$ gauge group of the SM. For the numerical analysis, we have considered all three types of charged leptons as well as six quarks, and therefore, $m_i=\{m_t,m_b,m_c,m_s,m_u,m_d,m_e,m_\mu,m_\tau\}$. 
Here, $m_t, m_b$ are the measured masses of the top and bottom quarks at $T=0$ and so on.

Within this regularization framework, the choice of the renormalization scale $\mu$ introduces an inherent uncertainty in the determination of the critical temperature and other related quantities~\cite{Chiang:2018gsn,Athron:2022jyi,Croon:2020cgk,Gould:2021oba}. To mitigate these uncertainties, several approaches have been employed for computing the Coleman-Weinberg (CW) potential, notably the on-shell (OS) and on-shell-like (OS-like) schemes. Among these, the OS-like scheme is extensively adopted in the literature, as it prescribes fixed renormalization scales for each particle species and maintains the tree-level parameter relations even at higher-loop levels, closely mirroring the behavior of the OS scheme. In this formalism, for instance, the CW potential takes the form:
%
%
\begin{equation} \label{}
V_{\rm 1-loop}^{\rm CW} (h,\phi_1,..\phi_n)
= \frac{1}{64\pi^2} \sum_{i} n_i \left\{  m_i^4(h,\phi_i) \left[ \log\left( \frac{m_i^2(h,\phi_i)}{m_i^2(v,0 )}\right)  - \frac{3}{2} \right] 
+ 2 m_i^2(h,\phi_i)m_i^2(v,0)  \right\},
\end{equation}
where $m_i^2(v,0)$ is the mass-squared of the species $i$ at zero temperature.  However, both OS schemes exhibit an infrared divergence arising from the Goldstone bosons, which become massless at zero temperature. Although this subtle issue can be addressed using the approaches outlined in Refs.~\cite{Martin:2014bca,Elias-Miro:2014pca}, our study primarily employs the $\overline{MS}$ scheme throughout. Next, we compute the thermal corrected potential.

 The one-loop potential at $T\neq 0$ is given as~\cite{Dolan:1973qd,Weinberg:1974hy}:
\begin{equation} \label{eq:finiteT}
\begin{split}
V^T_{\rm 1-loop} (h, \phi_1,..\phi_n, T)
&= \frac{T^4}{2\pi^2} \left[ \sum_{B} n_B J_B \left(\frac{m_B^2(h,\phi_i)} {T^2}\right) + \sum_{F} n_F J_F \left(\frac{m_F^2(h,\phi_i)} {T^2}\right)  \right],
\end{split}
\end{equation}
Here, B stands for all bosonic degrees of freedom that couple directly to the Higgs boson; therefore, B = $\{W, Z, \phi_1,.. \phi_n, \chi_{1,2,3}\}$ and F stands for three charged leptons and six quarks as mentioned earlier. $J_B$ and $J_F$ are the thermal bosonic and fermionic functions, which are as follows.
\begin{eqnarray}
J_{B/F} \left(\frac{m^2(h,\phi_i)} {T^2}\right)& = & \int_0^\infty dx x^2 \log\left( 1 \mp e^{-\sqrt{x^2+\frac{m^2(h,\phi_i)} {T^2}}}\right) 
\end{eqnarray}
For $T^2\gg m^2$, thermal bosonic and fermionic functions take an expansion as~\cite{Quiros:1999jp}
\begin{eqnarray} \label{eq:expn1}
&& J_B \left[\frac{m_B^2}{T^2} \right] = -\frac{\pi^4}{45} + \frac{\pi^2}{12T^2} m_B^2 - \frac{\pi}{6T^3} (m_B^2)^{3/2} - \frac{1}{32T^4}m_B^4 \ln \frac{m_B^2}{a_b T^2} + \cdots,   \\ \label{eq:expn2}
&&J_F \left[\frac{m_F^2}{T^2} \right] =  \frac{7\pi^4 }{360} - \frac{\pi^2}{24T^2} m_F^2 - \frac{1}{32T^4}m_F^4 \ln \frac{m_F^2}{a_f T^2} + \cdots, 
\end{eqnarray}
The high-temperature expansion with logarithmic terms, where $a_f = \pi^2 e^{3/2 - 2\gamma_E}$, $a_b = 16\pi^2 e^{3/2 - 2\gamma_E}$, and $\gamma_E = 0.577$, is accurate within $5\%$ for $y = \sqrt{m^2/T^2} \leq 1.6$ (fermions) and $y \leq 2.2$ (bosons)~\cite{Laine:2016hma,Anderson:1991zb}, and agrees with exact results to within $10\%$ for $y \leq 1\!-\!3$~\cite{Curtin:2016urg}.

\paragraph{}
Here, we also need to consider the contribution of the ring diagram to take care of infrared divergences at finite temperatures. This contribution can be given by:
\begin{equation}\label{eq:ring}
V_{\rm ring}(h ,\phi_1,..\phi_n, T)=-\frac{T}{12\pi}n_i\Big[\big(m_i^2(h, \phi_1,..\phi_n, T)\big)^{3/2}-\big(m_i^2(h, \phi_1,..\phi_n)\big)^{3/2}\Big],
\end{equation}
where $m_i^2(h, \phi_1,..\phi_n, T)$ are thermal masses called Debye masses which can be given as~\cite{Weinberg:1974hy} 
\begin{equation}\label{eq:massT}
m_i^2(h, \phi_1,..\phi_n) \rightarrow m_i^2(h, \phi_1,..\phi_n, T) = m_i^2(h, \phi_1,..\phi_n)+  \Pi_iT^2 ,
\end{equation}
where the temperature-dependent self-energy contribution $\Pi_i$ corresponding to the one-loop resumed diagrams to leading powers of the temperature from bosons and fermions is discussed in the next subsection.
\subsection*{Thermal self-energy corrections}
From the high temperature expansion of the thermal functions, the self-energy correction to the thermal masses of the CP-even scalar can be written as
\begin{align}\label{eq:daisy}
&\Pi_{h h} = \frac{3g^2}{16}  + \frac{g^{\prime 2}}{16}  + \frac{y_q^2}{4} + \frac{y_\ell^2}{12} + \frac{\lambda_H}{2} + \sum_i \frac{\lambda_{H\phi_i}}{24}
,  \\
&\Pi_{\phi_i\phi_i} = \frac{\lambda_{\phi_i}}{4} + \frac{\lambda_{H\phi_i}}{6}+ \sum_j \frac{\lambda_{ij}}{24}  , \\
&\Pi_{h \phi_i} = \frac{\lambda_{H\phi_i}}{12},\quad \Pi_{\phi_i \phi_j} = \frac{\lambda_{ij}}{12}.\label{eq:daisy1}
\end{align}
The thermal re-summation masses of the CP-odd Goldstone bosons $\chi_i$ are
\begin{eqnarray}
    \Pi_{\chi_i} = \frac{3g^2}{16}  + \frac{g^{\prime2}}{16}  + \frac{y_q^2}{4} +\frac{y_\ell^2}{12} + \frac{\lambda_H}{2} + \sum_i \frac{\lambda_{H\phi_i}}{24}.
\end{eqnarray}
Therefore, unlike $T=0$, the Higgs boson and singlet scalars mix at higher temperatures because of their non-zero $vev$s. The temperature-dependent $(n+1)\times (n+1)$ symmetric mass-square matrix for the CP-even scalars, in terms of classical background fields, is given as
\begin{equation}\label{mat:massT}
  M^2(h,\phi_1,..\phi_n,T)  =
  \begin{pmatrix}
      m_{hh}^2 & m_{h\phi_1}^2 &...m_{h\phi_n}^2\\ 
      m_{h\phi_1}^2 & m_{\phi_1\phi_1}^2 &...m_{\phi_1\phi_n}^2\\ 
    .......&........&.........\\
      m_{h\phi_n}^2 & m_{\phi_1\phi_n}^2 &...m_{\phi_n\phi_n}^2\\ 
  \end{pmatrix}+
  T^2\begin{pmatrix}
      \Pi_{hh} & \Pi_{h\phi_1} &...\Pi_{h\phi_n}\\ 
      \Pi_{h\phi_1} & \Pi_{\phi_1\phi_1} &...\Pi_{\phi_1\phi_n}\\ 
      .....&...&...\\
      \Pi_{h\phi_n} & \Pi_{\phi_1\phi_n} &...\Pi_{\phi_n\phi_n}\\ 
  \end{pmatrix},
\end{equation}
where the field-dependent mass-square matrix elements are given in Eqs.~\eqref{eq:masses}-\eqref{eq:masses1} and the daisy contributions are given in Eqs.~\eqref{eq:daisy}-\eqref{eq:daisy1}.
On the other hand, the longitudinal polarization states of $W$ bosons get thermal corrected masses~\cite{Oikonomou:2024jms}:
\begin{align}
   & m_{W_L}^2(h) \rightarrow m_{W_L}^2(h,T)= m_{W_L}^2 (h)+ \Pi_{W_L}(T), \quad \Pi_{W_L}(T) = \frac{11}{6}g^2 T^2.
\end{align}
At $T\neq 0$, there is a mixing between the longitudinal modes of $Z$ boson and $\gamma$, and the mixing mass matrix is given by~\cite{Oikonomou:2024jms}
\begin{equation}\label{eq:matrixZ}
m_{Z_L/\gamma_L}^2(h,T) =
    \begin{pmatrix}
    \frac{1}{4}g^2 h^2 + \frac{11}{6}g^2 T^2 & -\frac{1}{4}g g' h^2\\
    -\frac{1}{4}g g' h^2 & \frac{1}{4}g'^2 h^2 + \frac{11}{6}g'^2 T^2\\
    \end{pmatrix} \, .
\end{equation}
The eigenvalues of the above mass matrix give the mass of the $Z_L$ and $\gamma_L$.
%
\begin{equation}\label{Photon-thermalmass}
    m^2_{Z_L/\gamma_L} = \frac{1}{2} \left[ \frac{g^2 + g^{\prime 2}}{4}  h^2 + \frac{11\left(g^2 + g^{\prime 2} \right)}{6}  T^2 \pm \sqrt{\left(g^2 - g^{\prime 2} \right)^2 \left( \frac{h^2}{4}  + \frac{11T^2}{6}  \right)^2  + \frac{g^2 g^{\prime 2}}{4} h^4 }\right].
\end{equation}
The resultant bosonic degrees of freedom are \{$\varphi_1,..\varphi_n, \chi_{i}, W_L, Z_L, W_T, Z_T,\gamma_L$\} whereas fermionic degrees of freedom are $\{t,b,c,s,u,d,\tau,\mu,e\}$. Here, $m_{\varphi_1}^2,..m_{\varphi_n}^2$ are the eigenvalues of the mass-squared matrix at high temperature given in Eq.~\eqref{mat:massT}.
The constant factor $C_i$ in the CW potential given in Eq.~\eqref{eq:CWpot} is $3/2$ for scalars, fermions, and the longitudinal components of gauge bosons, whereas for the transverse modes of the gauge bosons, $C_i = 1/2$~\cite{Wainwright:2011kj}.

We introduce a counter-term $V_{\rm CT}(h,\phi_1,..\phi_n, T)$ to the effective potential in order to ensure that the zero-temperature masses of all bosonic and fermionic degrees of freedom, as well as the $vev$ of the Higgs field, remain equal to those determined from the tree-level potential. After including all contributions, the one-loop effective potential at $T \neq 0$~\cite{Quiros:1999jp,Espinosa:1993bs} is given by:
\begin{eqnarray}\label{eq:effpot}
    V_{\rm eff}(h,\phi_1,..\phi_n,T) &=& V_0(h,\phi_1,..\phi_n) + V_{\rm 1-loop}^{\rm CW}(h,\phi_1,..\phi_n,T) + V_{\rm 1-loop}^T(h,\phi_1,..\phi_n,T) \nonumber\\
    &+& V_{\rm CT}(h,\phi_1,..\phi_n,T)
\end{eqnarray}
The expression of counter-term potential $V_{\rm CT}(h,\phi_1,..\phi_n,T)$ is given in App.\,\ref{app:counter-term}. To take daisy contribution either we can add $V_{\rm ring}(h, \phi_1,..\phi_n, T)$ given in Eq.~\eqref{eq:ring} to the effective potential or we replace the field dependent masses $m_i^2(h, \phi_1,..\phi_n)$ of the bosons with their resumed masses $m_i^2(h, \phi_1,..\phi_n, T)$ in the one-loop CW potential $V_{\rm 1-loop}^{\rm CW}(h, \phi_1,..\phi_n, T)$ and finite temperature corrected potential $V_{\rm 1-loop}^T(h, \phi_1,..\phi_n, T)$ as discussed in Ref.~\cite{Parwani:1991gq}. 
\paragraph{}
It should be noted that the potentials $V_{\rm 1\text{-}loop}^{\rm CW}(h, \phi_1, \ldots, \phi_n, T)$ and $V_{\rm CT}^T(h, \phi_1, \ldots, \phi_n, T)$ depend explicitly on the temperature due to the thermal resummation of masses arising from the daisy diagrams. This resummation is essential to maintain the validity of the perturbative expansion at the one-loop level. Moreover, every term on the right-hand side of Eq.~\eqref{eq:effpot} also carries an implicit temperature dependence through the $vevs$ of the Higgs field and the singlet scalars in a certain temperature range. 

We now turn to the analysis of the effective potential, paying special attention to its minima at finite temperature. The critical temperature $T_c$ of the phase transition is defined as the temperature at which the minima of the potential becomes degenerate.
For a given parameter space, the conditions for determining the critical temperature $T_c$, the Higgs $vev$ at $T_c$, denoted $v_c$, and the $vev$ of each singlet scalar field $\phi_{ic}$ at $T_c$ are given by

\begin{eqnarray}
  &&  \frac{d V_{\rm eff}(h,\phi_1,..\phi_n,T_c)}{dh}\Big|\phi_{\rm low}=0,\quad \frac{d V_{\rm eff}(h,\phi_1,..\phi_n,T_c)}{d\phi_i}\Big|\phi_{\rm high}=0,\\
&&     V_{\rm eff}(\phi_{\rm low},T_c) = V_{\rm eff}(\phi_{\rm high},T_c),
\end{eqnarray}
where $\phi_{\rm low}$ and $\phi_{\rm high}$ are defined by $\phi_{\rm low}\equiv (h=v_c,\phi_i\approx 0)$ and $\phi_{\rm high}\equiv (h\approx 0,\phi_i=\phi_{ic})$.
The phase transition strength along the field's direction is defined by 
\begin{eqnarray}
    \xi_H = \frac{v_c}{T_c},\quad \xi_i=\frac{\phi_{ic}}{T_c}.
\end{eqnarray}
Basically, the strength of the phase transition is numerically characterized by the ratio of the difference in high $vevs$ and low $vevs$ to the critical temperature $T_c$ which are provided in Tab.\,\ref{tab:FOPT2} and \ref{tab:FOPT3} for two and three singlet scenarios.
The necessary condition for the strong first-order electroweak phase transition (SFOEWPT) is $\xi_H\ge 1$ along the direction of the Higgs field.
In this work, we have used the cosmological phase transition code \texttt{CosmoTransition} package~\cite{Wainwright:2011kj} to determine the phase transition pattern and to calculate the critical temperature $T_c$ and the $vevs$ of the scalar fields at $T_c$. We have employed the $\overline{MS}$ renormalization scheme, where the one-loop Coleman-Weinberg (CW) potential explicitly depends on the choice of the renormalization scale $\mu$ (see Eq.~\eqref{eq:CWpot}). To reduce this scale dependence, a renormalization group equation (RGE) improvement of the one-loop CW potential should be performed, as discussed in Refs~\cite{Andreassen:2014eha,Andreassen:2014gha}. However, this implementation is left for future work. In the present analysis, we take the mass of the top quark as the renormalization scale of the theory.

\begin{figure}
    \centering
    \includegraphics[width=0.49\linewidth]{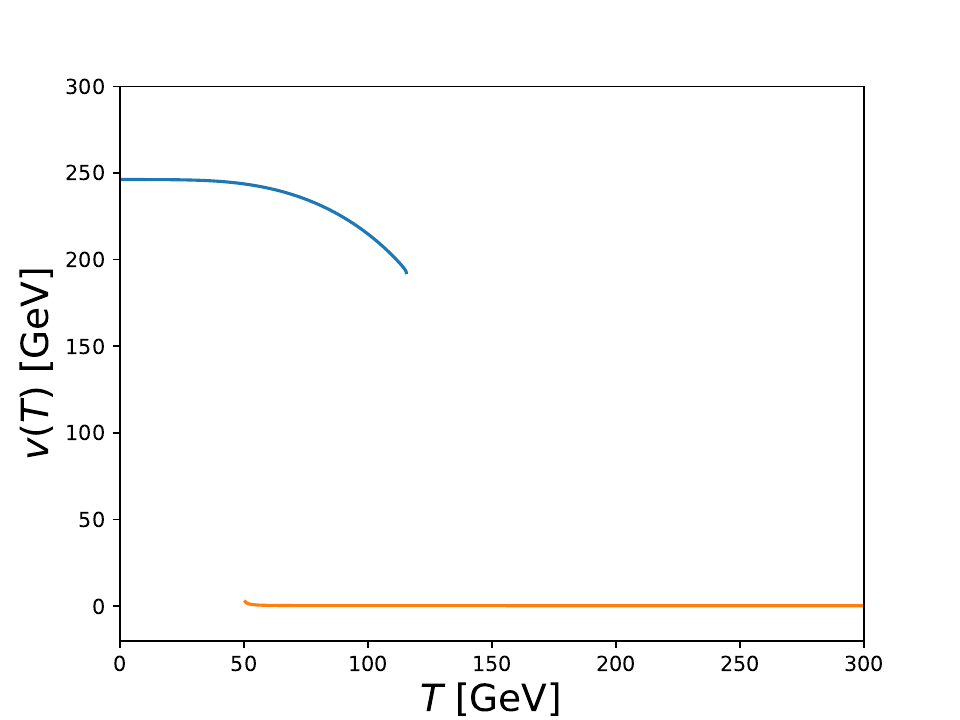}
    \includegraphics[width=0.49\linewidth]{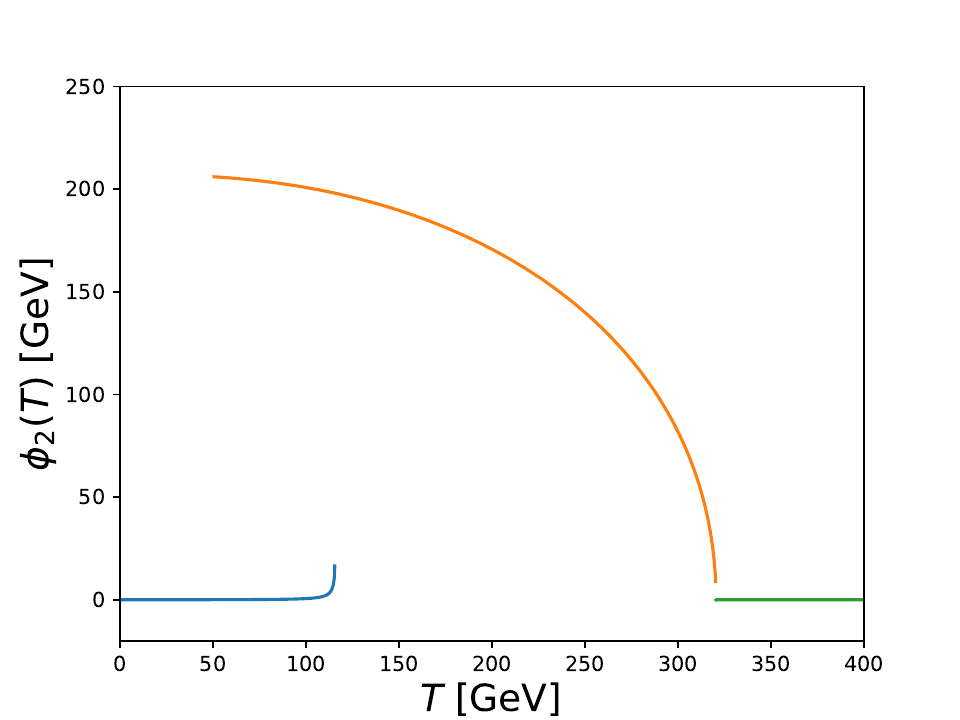}
    \caption{Variation of $vev$ of the Higgs field (left) and singlet scalar field $\phi_2$ (right) with temperature for BP1 for two-singlet extensions.}
    \label{fig:vev_two}
\end{figure}

The pattern of the first-order EWPT in the two-scalar scenario is illustrated in Fig.\,\ref{fig:vev_two}, where the evolution of the vacuum expectation values ($vevs$) of the Higgs field (left panel) and the singlet scalar field $\phi_2$ (right panel) are shown for benchmark point BP-1. Similar phase transition behavior is observed for BP-2 and BP-3.

At very high temperatures, all scalar fields, including the Higgs and singlets, have vanishing $vevs$, indicating that the universe was in a fully symmetric phase. As the temperature decreases, the singlet scalar field $\phi_2$ acquires a non-zero $vev$ due to a sizable Higgs portal coupling, while the Higgs field remains without a $vev$. In contrast, the other singlet scalar $\phi_1$ remains in the symmetric phase throughout the thermal evolution of the universe because of its tiny Higgs portal coupling.

Consequently, at this intermediate temperature range, the discrete symmetry $\mathcal{G}$ is spontaneously broken via the $vev$ of $\phi_2$, while electroweak symmetry remains intact. As the temperature drops further, the Higgs field eventually develops a $vev$, triggering the spontaneous breaking of electroweak symmetry. Simultaneously, the $vev$ of $\phi_2$ vanishes, leading to the restoration of the $\mathcal{G}$ symmetry.
This sequential symmetry-breaking pattern demonstrates a non-trivial two-step phase transition: first breaking $\mathcal{G}$ at higher temperatures, followed by electroweak symmetry breaking at lower temperatures, characteristic of a strongly first-order EWPT.

We summarize phase transition characteristics and relevant benchmark values, including nucleation temperature, vacuum expectation values, and gravitational wave parameters in Tabs.\,\ref{tab:FOPT2} and \ref{tab:FOPT3} for the two- and three-scalar scenarios, respectively~\footnote{These $vevs$ are reported as integers in units of GeV for clarity. Additionally, all benchmark values of \(\alpha_N\) and the strength parameters \((\xi_H, \xi_i)\) are rounded to two decimal places to maintain consistency and precision in presentation.}.

From Tab.\,\ref{tab:FOPT2}, we see that a strong first-order phase transition occurs along both the Higgs and the $\phi_2$ directions for all three benchmark points. The electroweak phase transition is strongest for BP-1 due to its larger Higgs portal coupling, $\lambda_{H\phi_2}$, and becomes progressively weaker for BP-2 and BP-3 as this coupling decreases at fixed mass. Therefore, to achieve a strong first-order electroweak phase transition, the Higgs portal coupling must be large enough for the singlet scalar to acquire a nonzero $vev$ at high temperature.

\begin{figure}
    \centering
    \includegraphics[width=0.49\linewidth]{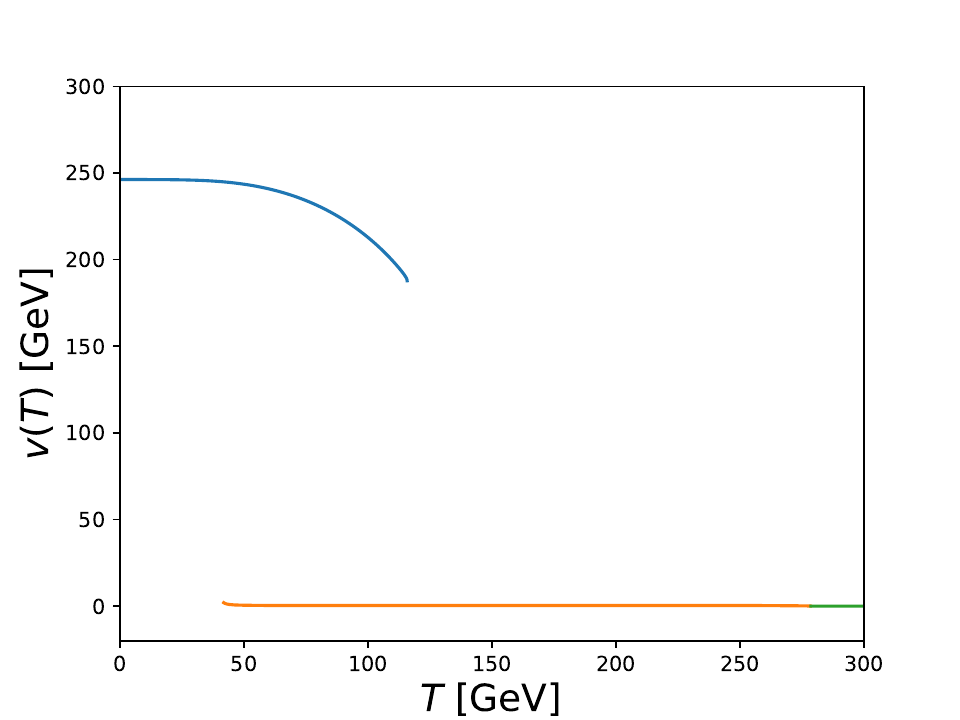}
    \includegraphics[width=0.49\linewidth]{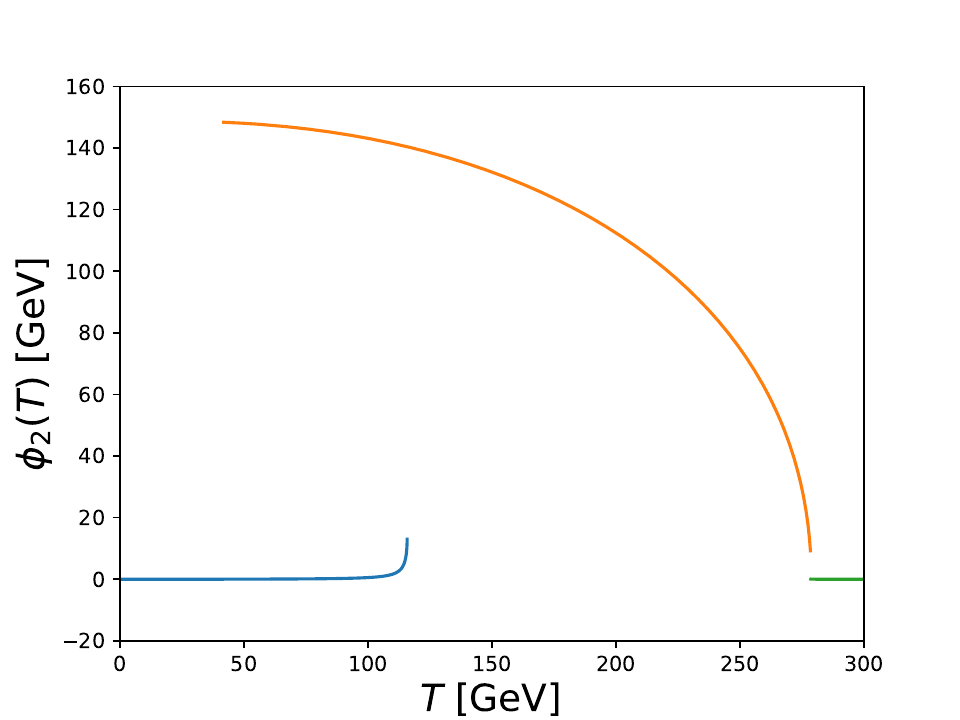}
    \caption{Variation of $vev$ of the Higgs field (left) and singlet scalar field $\phi_2$ (right) with temperature for BP1 for three-singlet extensions. Note that the singlet scalar field $\phi_3$ follows the same pattern as $\phi_2$. }
    \label{fig:vev_three}
\end{figure}

In the three-singlet scalar scenario, we introduce an additional scalar field, $\phi_3$, which is a copy of $\phi_2$. The corresponding phase transition strength parameters are given in Tab.\,\ref{tab:FOPT3}. Fig.\,\ref{fig:vev_three} shows the temperature evolution of the $vevs$ for the Higgs field (left panel) and the singlet scalar $\phi_2$ (right panel). The evolution of the $vev$ for $\phi_3$ follows the same phase transition pattern as $\phi_2$, as shown in the right panel of Fig.~\ref{fig:vev_three}. As in the two-singlet scenario, the singlet scalar field $\phi_1$ maintains a vanishing $vev$ throughout the thermal history of the universe for sufficiently small values of its Higgs portal coupling. It is important to note that the quartic couplings $\lambda_{\phi_i}$, as well as the mixed self-interaction terms $\lambda_{\phi_i\phi_j}$, play a crucial role in determining the nature of the phase transition in both the two-scalar and three-scalar scenarios.

A comparison of the benchmark points in Tab.\,\ref{tab:FOPT2} and~\ref{tab:FOPT3} shows that the three-singlet scenario leads to a stronger phase transition strength along the Higgs direction. This enhancement arises from the additional contribution of $\phi_3$ to the effective potential. Since $\phi_3$ has the same mass and Higgs portal coupling as $\phi_2$, it contributes equally to strengthening the phase transition.

\begin{table}[htbp]
\centering
\renewcommand{\arraystretch}{1.3}
\begin{tabular}{|c|c|c|c|}
\hline
& \multicolumn{3}{|c|}{\textbf{Two-scalar scenario} } \\
\hline
\textbf{PT parameters} &  BP1 & BP2 & BP3 \\
\hline
\shortstack{ $(v(T), \phi_{i}(T))|_{T_c}$\\ (high $vevs$)}  & $(0,0,202)$  & $(0,0,205)$ & $(0,0,188)$  \\
\hline
\shortstack{ $(v(T), \phi_{i}(T))|_{T_c}$\\ (low $vevs$)}  & $(222,0,0)$  & $(200,0,0)$  & $(174,0,1)$  \\
\hline
$T_c$ (GeV) &  $92$ & $112$ & $127$  \\
\hline
 $(\xi_H,\xi_i)$ &$(2.41,0,2.20)$ &$(1.79,0,1.83)$ &$(1.37,0,1.47)$\\
\hline
\shortstack{ $(v(T), \phi_{i}(T))|_{T_N}$\\ (high $vevs$)}  & $(0,0,205)$  & $(0,0,208)$ & $(2,0,190)$  \\
\hline
\shortstack{ $(v(T), \phi_{i}(T))|_{T_N}$\\ (low $vevs$)}  & $(241,0,0)$ &  $(219,0,0)$& $(189,0,0)$ \\
\hline
$T_N$ (GeV)   & $60$ & $97$ & $120$\\
\hline
$\alpha_N$  &0.16 & 0.04 &0.02\\
\hline
$\beta/H_N$  & $301$ & $1770$& $5834$\\
\hline
\end{tabular}
\caption{The values of phase transition strength quantities ($\xi_H,\xi_i$) and characteristic parameters for GWs for two-singlet extensions. Here, $vevs$ $(v(T),\phi_i(T))$ are in GeV unit.  }
    \label{tab:FOPT2}
\end{table}

\begin{table}[htbp]
\centering
\renewcommand{\arraystretch}{1.3}
\begin{tabular}{|c|c|c|c|}
\hline
& \multicolumn{3}{|c|}{\textbf{Three-scalar scenario} } \\
\hline
\textbf{PT parameters} &  BP1 & BP2 & BP3 \\
\hline
\shortstack{ $(v(T), \phi_{i}(T))|_{T_c}$\\ (high $vevs$)}  & $(0,0,144,144)$  & $(0,0,148,148)$ & $(0,0,138,138)$  \\
\hline
\shortstack{$(v(T), \phi_{i}(T))|_{T_c}$\\ (low $vevs$)} & $(222,0,0,0)$  & $(200,0,0,0)$  & $(176,0,0,0)$  \\
\hline
$T_c$ (GeV)&  $91$ & $111$ & $125$  \\
\hline
 $(\xi_H,\xi_i)$ &$(2.44,0,1.58.1.58)$ &$(1.80,0,1.33,1.33)$ &$(1.41,0,1.10,1.10)$\\
\hline
\shortstack{$(v(T), \phi_{i}(T))|_{T_N}$\\(high $vevs$)}  & $(0,0,148,148)$  & $(1,0,151,151)$  & $(1,0,140,140)$  \\
\hline
\shortstack{$(v(T), \phi_{i}(T))|_{T_N}$\\(low $vevs$)}  & $(243,0,0,0)$ &  $(222,0,0,0)$ & $(194,0,0,0)$ \\
\hline
$T_N$ (GeV)  & $51$ & $92$ & $115$\\
\hline
$\alpha_N$  &0.25 & 0.05 &0.02\\
\hline
$\beta/H_N$  & $412$ & $1501$& $4405$\\
\hline
\end{tabular}
\caption{Same caption as Tab.\,\ref{tab:FOPT2}, but this is for three-singlet extensions to the SM. }
    \label{tab:FOPT3}
\end{table}

\section{Stochastic Gravitational Waves}
\label{sec:gw}
In the early universe, a first-order phase transition (FOPT) can act as a significant source of stochastic gravitational waves (GWs). This transition typically occurs around \( \sim 10^{-11}~\text{s} \) after the Big Bang, prior to the epoch of Big Bang Nucleosynthesis (BBN). As the universe cools down, it undergoes a phase transition at a critical temperature \( T_c \). Above \( T_c \), the universe exists in a symmetric phase. However, as the temperature drops below \( T_c \), a second, degenerate vacuum state appears in the potential, initiating the dynamics of the FOPT.

During this transition, the release of latent heat plays a crucial role. A fraction of this energy contributes to the production of gravitational waves, while the remaining part is absorbed by the surrounding plasma. The evolution of a FOPT is governed by the process of bubble nucleation. Consequently, determining the nucleation temperature is essential for calculating key phenomenological parameters, which are needed to estimate the resulting gravitational wave spectrum.

As the temperature falls below \( T_c \), the probability of quantum tunneling from the false vacuum to the true vacuum becomes significant. The tunneling rate per unit volume, \( \Gamma(T) \), is expressed as follows~\cite{Grojean:2006bp,Linde:1981zj}:
\begin{equation}\label{eq:tunnelling-prob}
\Gamma(T) \approx T^4 \left( \frac{S_E(T)}{2\pi T} \right)^{3/2} e^{-\frac{S_E(T)}{T}},
\end{equation}
where \( S_E (T)\) represents the three-dimensional Euclidean action (also known as the bounce action). This action is a measure of the tunneling process and is given by~\cite{Linde:1981zj}:
\begin{equation}\label{eq:bounce-action}
S_E(T) = \int_{0}^{\infty} 4\pi r^2 dr \left[ V_{\rm eff}(\phi, T) + \frac{1}{2} \left( \frac{d\phi(r)}{dr} \right)^2 \right].
\end{equation}
Here, \( \phi \) is the scalar field(s) that drives the phase transition, and its profile \( \phi(r) \) is determined as a function of the radial coordinate \( r \).  Here, \( V_{\rm eff}(\phi, T) \) is the finite temperature one-loop effective potential given in Eq.~\eqref{eq:effpot}. The scalar field satisfies the following equation of motion:
\begin{equation}\label{eq:phivariationr}
\frac{d^2 \phi}{dr^2} + \frac{2}{r} \frac{d\phi}{dr} = \frac{dV_{\rm eff}(\phi, T)}{d\phi}.
\end{equation}
This equation must be solved subject to the appropriate boundary conditions~\cite{Affleck:1980ac, Linde:1977mm}. The required conditions are:
\begin{equation}\label{bc:1}
\frac{d\phi}{dr} = 0 \quad \text{for} \quad r \to 0,
\end{equation}
and
\begin{equation}\label{bc:2}
\phi(r) \to \phi_{\text{false}} \quad \text{for} \quad r \to \infty.
\end{equation}

These equations and boundary conditions together determine the dynamics of the scalar field and the formation of the bubble, which are pivotal for understanding the characteristics of the FOPT and the associated production of stochastic gravitational waves. Further details on these computations can be found in \cite{Affleck:1980ac, Linde:1977mm}.

 In the parameter space considered, the intermediate phase associated with the spontaneous breaking of the discrete symmetry is short-lived and slightly biased by finite-temperature effects from the Higgs–portal interactions. This small asymmetry removes the vacuum degeneracy and causes any transient domain-wall network~\cite{Blasi:2022woz,Agrawal:2023cgp} to decay well before the second electroweak step, leaving negligible impact on the subsequent phase transition or the resulting gravitational-wave spectrum.

The term \( \phi_{\text{false}} \) in Eq.~\eqref{bc:2} refers to the scalar field value at the false vacuum in four-dimensional spacetime. To solve the equation of motion given in Eq.~\eqref{eq:phivariationr}, the \texttt {CosmoTransitions} package~\cite{Wainwright:2011kj} has been employed. This computational tool specializes in the analysis of cosmological phase transitions.

Two fundamental parameters, crucial for estimating the gravitational wave spectrum generated by first-order phase transitions, are the relative change in energy density during the transition, denoted by \( \alpha_N \), and the inverse duration of the phase transition, represented by \( \beta \). Both parameters are defined at the nucleation temperature \( T_N \).

The parameter \( \alpha_N \) is calculated as the ratio of the latent heat released during the phase transition to the radiation energy density:
\begin{equation}\label{eq:alphaP}
\alpha_N = \frac{\Delta \rho}{\rho_{\text{rad}}},
\end{equation}
where \( \rho_{\text{rad}} (= \pi^2 g_\ast T_N^4/30)\) is the radiation energy density, and \( \Delta \rho \) is the latent heat released during the transition. The latent heat \( \Delta \rho \) is determined as \cite{Kamionkowski:1993fg, Kehayias:2009tn}:
\begin{equation}\label{eq:delrho}
\Delta \rho = \left[ V_{\rm eff}(\phi_0, T) - T \frac{dV_{\rm eff}(\phi_0, T)}{dT} \right]_{T = T_N} 
- \left[ V_{\rm eff}(\phi_N, T) - T \frac{dV_{\rm eff}(\phi_N, T)}{dT} \right]_{T = T_N}.
\end{equation}
The corresponding values of \(\phi_0\) and \(\phi_N\) at the nucleation temperature \(T_N\) are listed in Tabs.\,\ref{tab:FOPT2} and \ref{tab:FOPT3} as low $vevs$ and high $vevs$ respectively. 

The parameter \( \beta \) characterizes the inverse of the phase transition duration relative to the Hubble time. It is defined as:
\begin{equation}\label{eq:betaP}
\frac{\beta}{H_N} = T_N \frac{d}{dT} \left( \frac{S_E(T)}{T} \right) \Bigg|_{T = T_{N}} ,
\end{equation}
where \( H_{N} \) is the Hubble expansion rate at the time of the phase transition, and \( T_{N} \) represents the nucleation temperature. The code calculates \( T_N \) by solving the equation \( S_E(T_N)/T_N= 140 \)~\cite{Wainwright:2011kj}. 

These parameters, \( \alpha_N \) and \( \beta \), are critical for understanding the dynamics of the phase transition and predicting the characteristics of the resulting stochastic gravitational wave background. 
A detailed summary of the nucleation temperature \(T_N\), the phase transition strength \(\alpha_N\), and the inverse time scale \(\beta/H_N\) is provided in Tabs.\,\ref{tab:FOPT2} and~\ref{tab:FOPT3}. We find that the three-singlet benchmarks (Tab. \ref{tab:FOPT3}) consistently exhibit \textit{lower nucleation temperatures} (e.g., \(T_N = 51~\mathrm{GeV}\) for BP1) and \textit{stronger transitions} (e.g., \(\alpha_N = 0.25\)), indicating enhanced supercooling. In contrast, the two-singlet cases (Tab.\,\ref{tab:FOPT2}) show higher \(T_N\) (e.g., \(T_N = 60~\mathrm{GeV}\) for BP1) and weaker transitions (e.g., \(\alpha_N = 0.16\)). Additionally, the three-singlet scenarios yield \textit{smaller values of \(\beta/H_N\)}, particularly in BP1, pointing to slower, more prolonged transitions. This feature is crucial, as it leads to stronger gravitational wave signals. Overall, these results highlight that extending the scalar sector to three singlets significantly enhances the strength and observability of gravitational waves from the electroweak phase transition, making such models highly promising for future GW experiments. Next, we estimate the gravitational wave energy density spectrum.

The GW energy density spectrum consists of three primary contributions: bubble wall collisions, sound waves, and magneto-hydrodynamic turbulence (MHD). These components collectively determine the total energy spectrum, which can be approximated as the sum of all three:
\begin{equation}\label{eq:GWTotal}
\Omega_{\rm GW}h^2 \approx \Omega_{\rm col} h^2 + \Omega_{\rm sw} h^2 + \Omega_{\rm tur} h^2,
\end{equation}
where the terms on the right-hand side represent the contributions from bubble wall collisions, sound waves, and MHD turbulence, respectively \cite{Caprini:2015zlo}. Here, \( h = H_0/(100 \, \text{km} \cdot \text{s}^{-1} \cdot \text{Mpc}^{-1}) \), where \( H_0 \) is the present-day Hubble constant \cite{DES:2017txv}.

The energy density contribution from bubble wall collisions can be derived using the envelope approximation \cite{Jinno:2016vai} and is expressed as a function of frequency \( f \):
\begin{equation}\label{eq:GWcoldetails}
\Omega_{\rm col} h^2 = 1.67 \times 10^{-5} {\left(\frac{\beta }{H_N} \right)^{-2}} \left( \frac{\kappa_c \alpha_N}{1 + \alpha_N} \right)^2 \left(  \frac{100}{g^{\ast}} \right)^{1/3} \left( \frac{0.11 v^3_w}{0.42 + v^2_w} \right) \frac{3.8 \left( f/f_{\rm col} \right)^{2.8}}{1 + 2.8 \left( f/f_{\rm col} \right)^{3.8}},
\end{equation}
where \( v_w \) is the bubble wall velocity, \( g_* \) is the effective degrees of freedom and for SM it is $106.75$~\cite{Husdal:2016haj}, and \( \kappa_c \) is the efficiency factor for bubble collisions, defined as \cite{Kamionkowski:1993fg,Borah:2023zsb}:
\begin{equation}\label{eq:kcfac}
\kappa_c = \frac{0.715 \alpha_N + \frac{4}{27} \sqrt{\frac{3 \alpha_N}{2}}}{1 + 0.715 \alpha_N}.
\end{equation}
The redshifted peak frequency \( f_{\rm col} \) is given by:
\begin{equation}\label{eq:PF1}
f_{\rm col} = 16.5 \times 10^{-6} \left( \frac{f_{\ast}}{\beta} \right) \left( \frac{\beta}{H_N} \right) \left( \frac{T_N}{100 \, {\rm GeV}} \right) \left( \frac{g^{\ast}}{100} \right)^{1/6} \, {\rm Hz},
\end{equation}
where the fitting function \( f_{\ast}/\beta \) is defined as \cite{Jinno:2016vai}:
\begin{equation}\label{eq:fastbybetadetails}
\frac{f_{\ast}}{\beta} = \frac{0.62}{1.8 - 0.1 v_w + v^2_w}.
\end{equation}
Assuming \( v_w = 1 \), as expanding bubbles achieve relativistic terminal velocities, simplifies the calculation.

The energy density contribution from sound waves can be expressed as \cite{Hindmarsh:2013xza,Hindmarsh:2016lnk,Hindmarsh:2017gnf}:
\begin{equation}\label{eq:GWswpart}
\Omega_{\rm sw} h^2 = 2.65 \times 10^{-6}\; \Upsilon(\tau_{\rm sw}) \left(  \frac{\beta}{H_N} \right)^{-1} v_w \left( \frac{\kappa_{\rm sw} \alpha_N}{1 + \alpha_N} \right)^2 \left( \frac{100}{g^{\ast}} \right)^{1/3} \left( \frac{f}{f_{\rm sw}} \right)^3 \left[ \frac{7}{4 + 3 \left( f/f_{\rm sw} \right)^2} \right]^{7/2},
\end{equation}
where \( \kappa_{\rm sw} \) is the efficiency factor indicating the fraction of latent heat converted into bulk plasma motion~\cite{Kamionkowski:1993fg,Borah:2025hpo}
\begin{equation}\label{eq:kappasw}
\kappa_{\rm sw} =\frac{\sqrt{\alpha_N}}{0.135+\sqrt{0.98+\alpha_N}}
\end{equation}
The peak frequency for sound waves is given by:
\begin{equation}\label{eq:PF2}
f_{\rm sw} = 1.9 \times 10^{-5} \left( \frac{1}{v_w} \right) \left( \frac{\beta}{H_N} \right) \left( \frac{T_N}{100 \, {\rm GeV}} \right) \left( \frac{g^{\ast}}{100} \right)^{1/6} \, {\rm Hz}.
\end{equation}
The factor \( \Upsilon(\tau_{\rm sw}) \), which accounts for the finite lifetime of sound waves, is defined as:
\begin{equation}\label{eq:swtimepart}
\Upsilon(\tau_{\rm sw}) = 1 - \frac{1}{\sqrt{1+2 \tau_{\rm sw} H_{\ast}}},
\end{equation}
where \( \tau_{\rm sw} \) is the sound wave lifetime. Following Ref.~\cite{Hindmarsh:2017gnf}, we write $\tau_{\rm sw}\approx R_N/\overline{U}_f$, where the mean bubble separation $R_N = (8\pi)^{1/3}v_w \beta_N^{-1}$ and the root-mean-squared fluid velocity $\overline{U}_f=\sqrt{3\kappa_{\rm sw}\alpha/4}$. 

Finally, the contribution from MHD turbulence, caused by the complete ionization of the plasma \cite{Caprini:2009yp}, is expressed as:
\begin{equation}\label{eq:GWturpart}
\Omega_{\rm tur} h^2 = 3.35 \times 10^{-4} \left( \frac{\beta}{H_N} \right)^{-1} v_w \left( \frac{\kappa_{\rm tur} \alpha_N}{1 + \alpha_N} \right)^{3/2} \left( \frac{100}{g^{\ast}}\right)^{1/3} \left[ \frac{\left( f/f_{\rm tur} \right)^3}{\left[ 1 + \left( f/f_{\rm tur} \right) \right]^{11/3} \left( 1 + \frac{8 \pi f}{h_{\ast}} \right)} \right],
\end{equation}
where \( h_{\ast} = 16.5\times 10^{-6} \left( \frac{T_N}{100 \, {\rm GeV}} \right) \left( \frac{g^{\ast}}{100} \right)^{1/6} \, {\rm Hz} \), the inverse Hubble time during GW production, redshifted to today. The peak frequency \( f_{\rm tur} \) is:
\begin{equation}\label{eq:PF3}
f_{\rm tur} = 2.7 \times 10^{-5} \frac{1}{v_w} \left( \frac{\beta}{H_N} \right) \left( \frac{T_N}{100 \, {\rm GeV}} \right) \left( \frac{g^{\ast}}{100} \right)^{1/6} \, {\rm Hz.}
\end{equation}
The turbulence efficiency factor \( \kappa_{\rm tur} \) is defined as \( \kappa_{\rm tur} = \epsilon \kappa_{\rm sw} \), where \( \epsilon \) represents the fraction of bulk motion converted into turbulence. Previous studies suggest \( \kappa_{\rm tur} \approx 0.1 \kappa_{\rm sw} \), a value adopted here for numerical calculations. 
\begin{figure}
    \centering
    \includegraphics[width=0.48\linewidth]{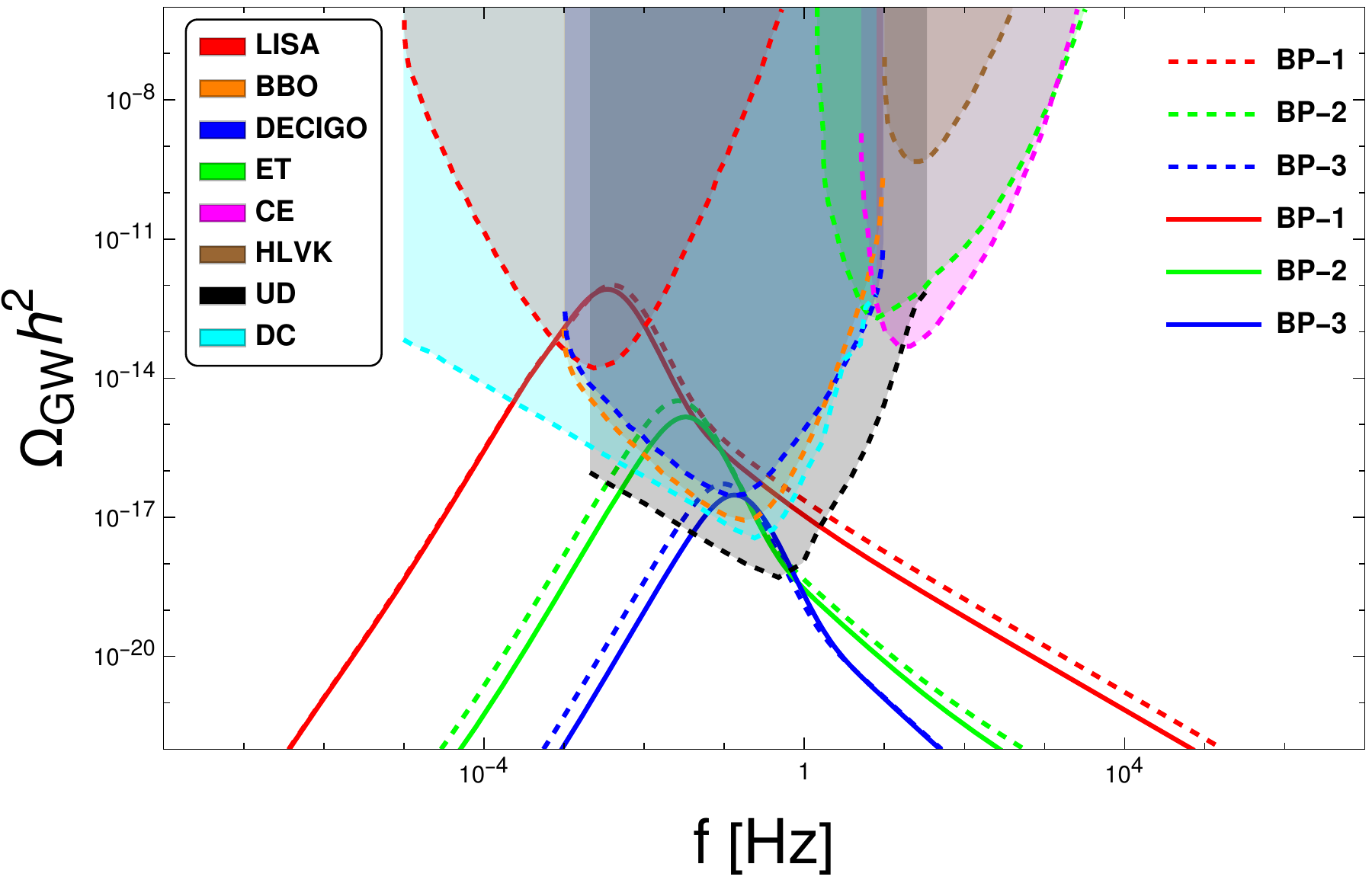}
    \includegraphics[width=0.48\linewidth]{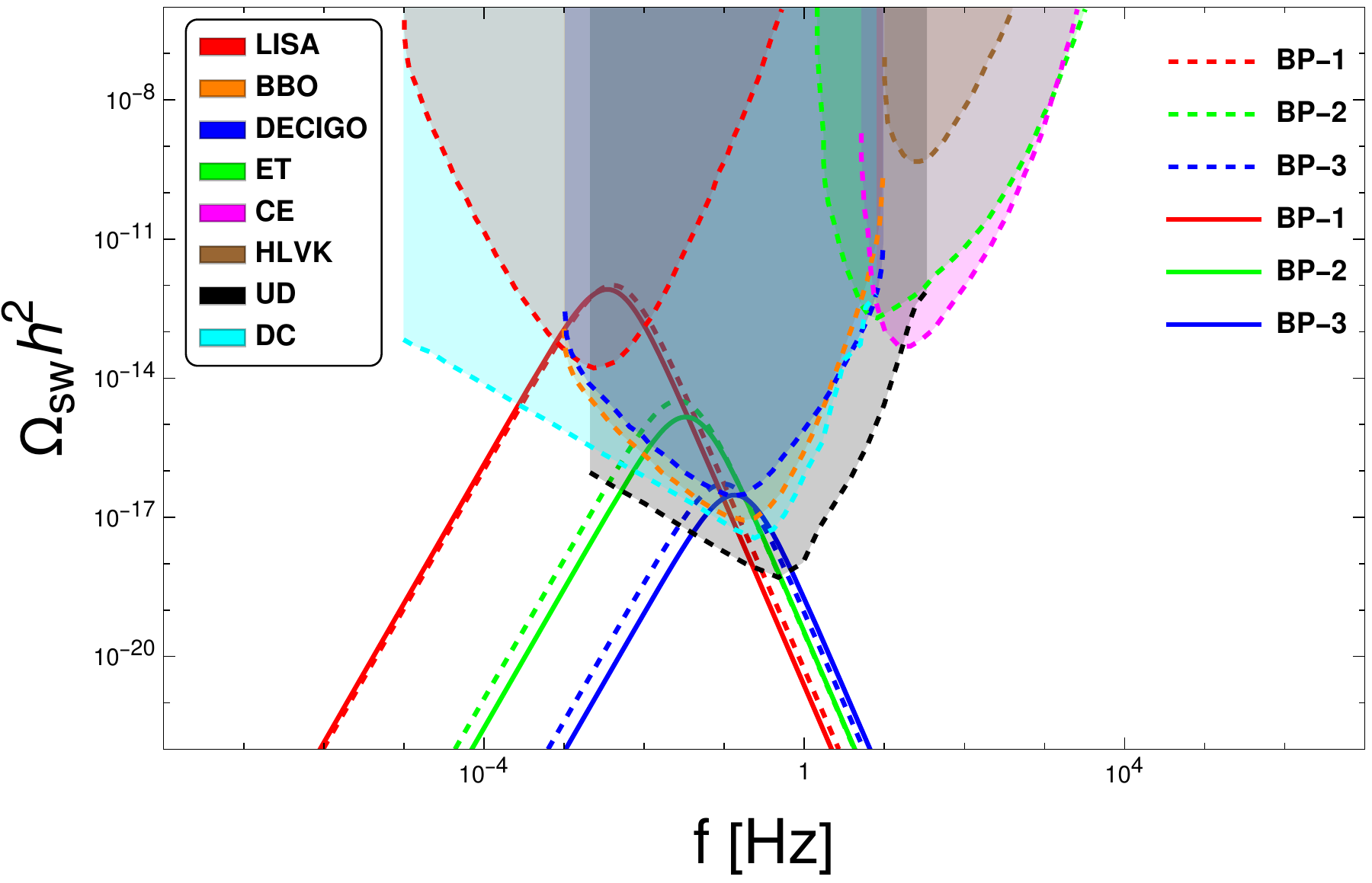}
    \includegraphics[width=0.48\linewidth]{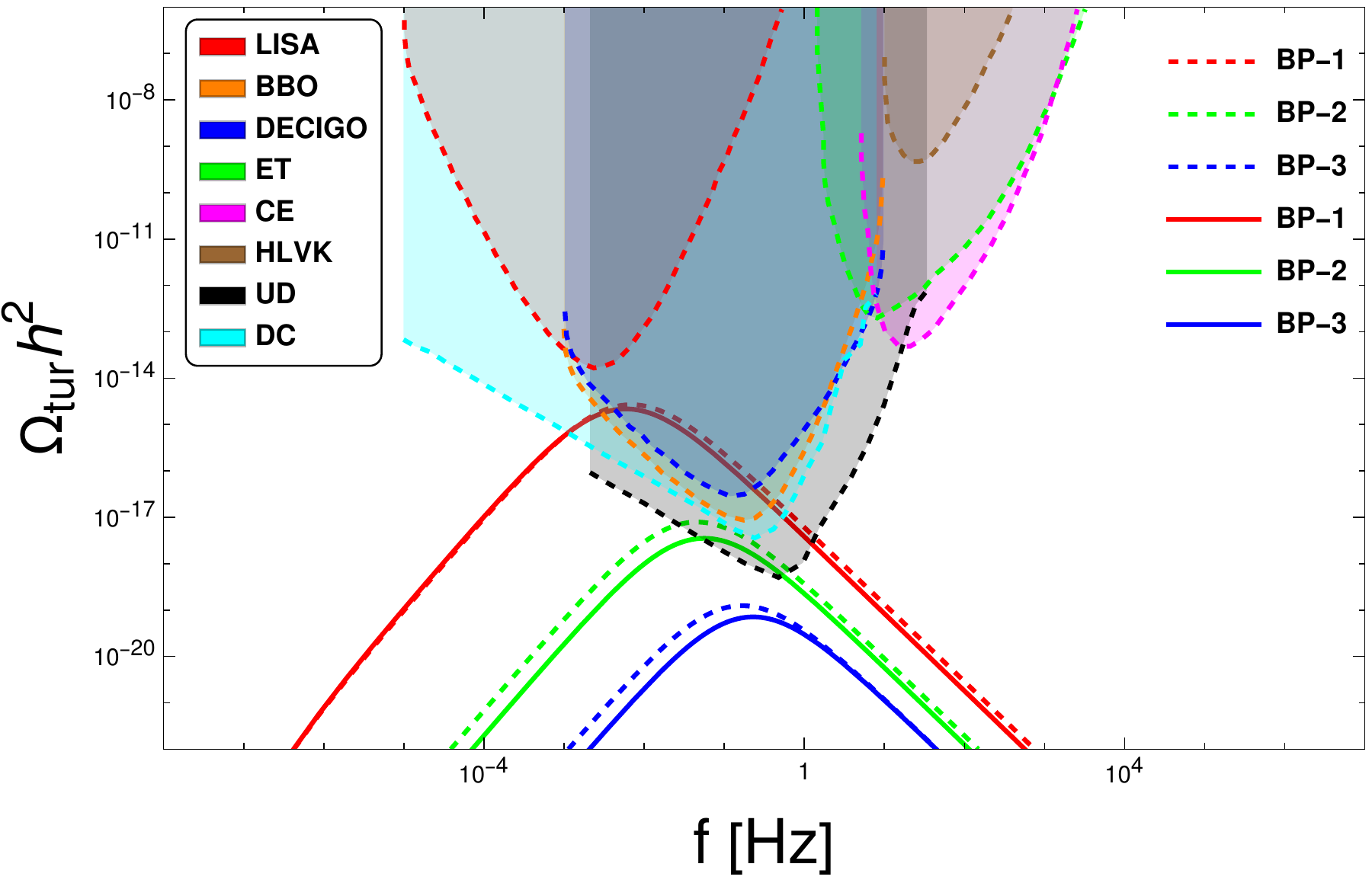}
    \includegraphics[width=0.48\linewidth]{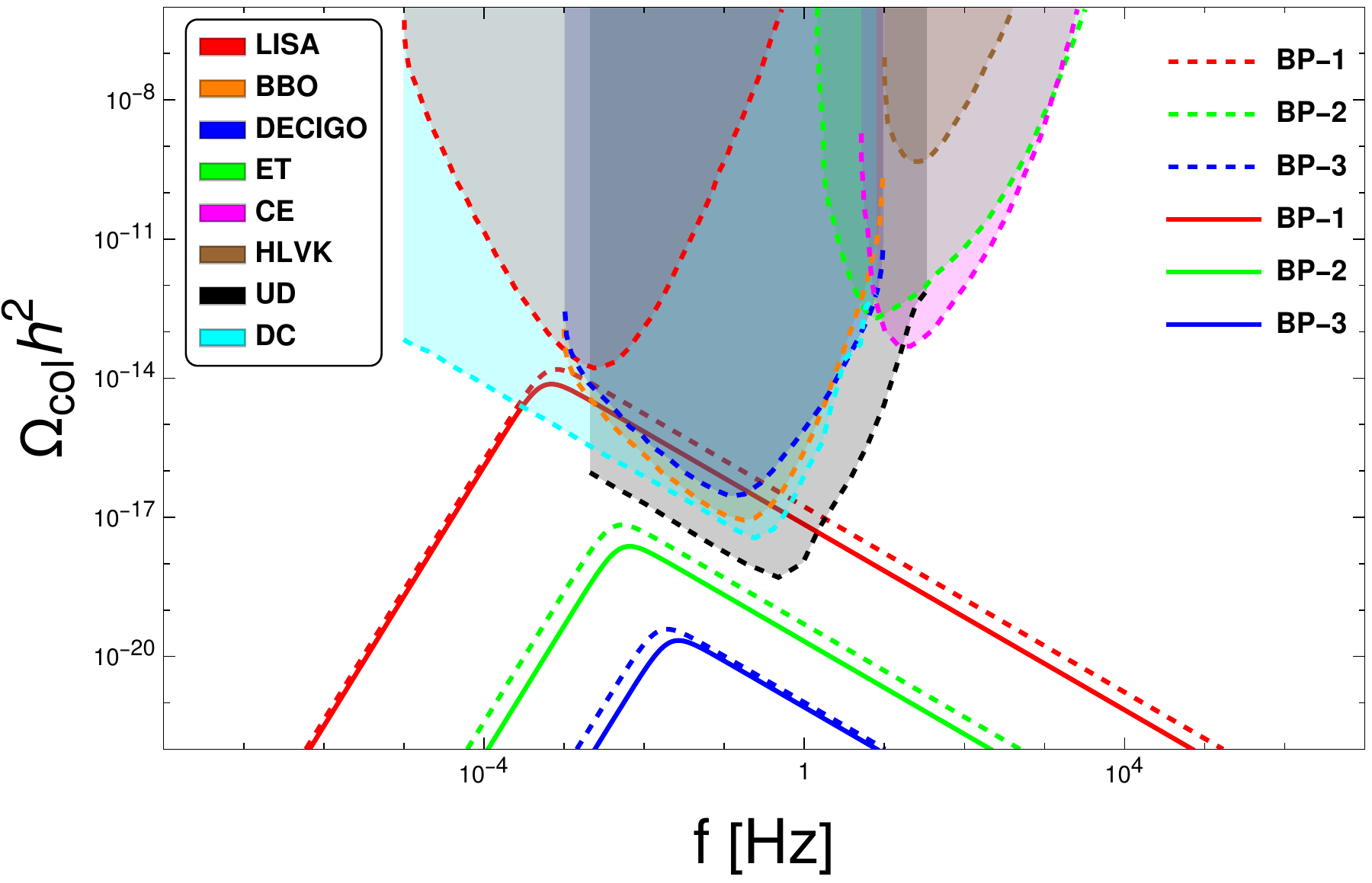}
\caption{
Gravitational wave spectra \( \Omega_{\rm GW} h^2 \) for benchmark points \texttt{BP-1}, \texttt{BP-2}, and \texttt{BP-3}, shown for both two-singlet (solid) and three-singlet (dashed) scenarios. The shaded regions represent the projected sensitivity curves of future gravitational wave interferometers—LISA, BBO, DECIGO, ET, CE, HLVK, ultimate-DECIGO(UD), and DECIGO-corr(DC)—as indicated by the legend.
}

    \label{fig:gwplot}
\end{figure}

Fig.~\ref{fig:gwplot} shows the energy-density spectrum $\Omega_{\rm GW} h^{2}$ of the stochastic gravitational waves generated during the electroweak phase transition for different benchmark points, plotted as a function of the frequency $f$. These spectra are compared against the projected sensitivities of present and upcoming gravitational-wave detectors, including LISA~\cite{LISA:2017pwj}, BBO~\cite{Yagi:2011wg}, DECIGO~\cite{Nakayama:2009ce}, Ultimate-DECIGO and correlated-DECIGO~\cite{Nakayama:2009ce}, ET~\cite{Punturo:2010zz}, CE~\cite{Reitze:2019iox}, and HLVK~\cite{LIGOScientific:2014pky,VIRGO:2014yos,KAGRA:2018plz}. The solid curves correspond to the two-singlet scalar scenario, whereas the dashed curves represent the three-singlet extension. The presence of additional singlets strengthens the electroweak phase transition by modifying the thermal effective potential, leading to larger latent-heat release, earlier bubble nucleation, faster bubble expansion, and consequently an enhanced gravitational-wave signal.

Figure~\ref{fig:gwplot} further decomposes the individual contributions to the total spectrum: the sound-wave component $\Omega_{\rm sw} h^{2}$ (top right), the turbulence contribution $\Omega_{\rm tur} h^{2}$ (bottom left), and the bubble-collision envelope component $\Omega_{\rm col} h^{2}$ (bottom right). In non-runaway regimes, friction between bubble walls and the surrounding plasma prevents relativistic runaway motion, causing long-lived acoustic waves to dominate the gravitational-wave production~\cite{RoperPol:2023dzg}. Among the benchmark points, \texttt{BP--1} exhibits the largest signal due to its strongest transition and highest value of $\alpha_{N}$, while \texttt{BP--3} yields the weakest signal. The peak frequencies reflect the underlying transition dynamics: \texttt{BP--1} peaks near $f \sim 0.01$~Hz and is accessible to LISA and DECIGO, \texttt{BP--2} peaks near $f \sim 0.1$~Hz and aligns well with DECIGO sensitivity, and \texttt{BP--3} peaks around $f \sim 1$~Hz, falling within the reach of future detectors such as DECIGO and BBO.

This frequency-dependent alignment between benchmark predictions and detector sensitivity curves highlights the complementarity of various gravitational wave observatories in probing extended scalar sector models that predict strong first-order electroweak phase transitions. Our analysis therefore demonstrates that such scenarios can be tested through upcoming GW detectors, offering a promising indirect avenue to explore new physics beyond the Standard Model.

\section{Summary and Conclusion}
\label{sec:conclusion}

The real scalar singlet extension of the Standard Model (SM) remains one of the most economical and well-motivated frameworks to account for dark matter (DM), relying on a discrete $\mathbb{Z}_2$ symmetry to stabilize the DM candidate. This minimal model, characterized by only two parameters—the DM mass and the Higgs portal coupling—has been extensively studied. However, the most recent results from direct detection experiments, such as LUX-ZEPLIN (LZ), have significantly constrained the viable parameter space. These constraints push the portal coupling to very small values, rendering the model difficult to probe through other experimental means. Moreover, the correct relic abundance is predominantly achieved near the Higgs resonance region, where $m_{\rm DM} \simeq m_h/2$ and with tiny Higgs portal coupling.

In this work, we revisit the scalar singlet DM scenario and extend the minimal setup by incorporating additional real singlet scalars, thereby enriching the dark sector. We investigate two extensions: one involving two singlet scalars, and another involving three singlet scalars. The presence of additional scalars—especially when one remains near the Higgs resonance to maintain the correct relic density—permits significantly larger Higgs portal couplings for the remaining scalar(s), while still satisfying all current experimental constraints.

We present three benchmark points for each scenario, all of which satisfy the observed relic abundance, are consistent with the latest spin-independent direct detection limits, comply with indirect detection constraints (notably gamma-ray line searches), and remain safe under collider bounds, including limits from invisible Higgs decays and LHC monojet searches with missing transverse momentum—prior to the electroweak phase transition (EWPT).

A key consequence of allowing larger portal couplings is their significant impact on the nature of the EWPT. In particular, they can induce a strong first-order electroweak phase transition (SFOPT), characterized by large values of the strength parameter $\alpha_N$, small values of the nucleation temperature $T_N$, and moderate values of $\beta/H_N$. Such phase transitions have potential to produce a stochastic gravitational wave (GW) background, potentially observable in upcoming GW experiments.

We compute the resulting gravitational wave spectra for all benchmark points and compare them with the projected sensitivities of future GW interferometers, including LISA, DECIGO, BBO, ET, CE, and HLVK. Notably, we find:
\begin{itemize}
    \item \textbf{BP-1} exhibits the strongest GW signal due to a large $\alpha_N$, peaking at $f \sim 0.01$ Hz, and is most accessible to LISA and DECIGO.
    \item \textbf{BP-2} peaks at $f \sim 0.1$ Hz, falling within the optimal range for DECIGO sensitivity.
    \item \textbf{BP-3} produces a moderate signal with peak frequency $f \sim 1$ Hz, potentially observable at BBO and DECIGO.
\end{itemize}

These results demonstrate that multi-singlet scalar extensions of the Higgs portal dark matter model not only open new regions of viable parameter space—otherwise excluded in the minimal setup—but also establish a direct connection between dark matter phenomenology, electroweak phase transition dynamics, and gravitational wave observables. As such, these scenarios can be probed in a complementary manner by future astrophysical, cosmological, and collider experiments.

\section*{Acknowledgment}
The authors acknowledge Dr. Abhijit Kumar Saha for the useful discussions. TS is supported by the National Natural Science Foundation of China under Grant Nos. 12475094, 12135006, and 12075097.
JD acknowledges the Indian Institute of Technology Kanpur (IITK) for the Institutional Postdoctoral Fellowship grant with File No. DF/PDF/2023-IITK/2183. JD also thanks SERB, Government of India, for the National Postdoctoral Fellowship (NPDF) grant with File No. PDF/2023/001540. The work of AC was supported by the Japan Society for the Promotion of Science (JSPS) as a part of the JSPS Postdoctoral Program (Standard), grant number: JP23KF0289. 
\appendix
\section{Counter-term potential}\label{app:counter-term}
 By adding $V_{\rm CT}(h,\phi_1,..\phi_n,T)$ to the effective potential, one can cancel the shifts in the Higgs mass and $vev$ induced by the zero-temperature one-loop Coleman–Weinberg (CW) contributions. When all singlet scalars are possible dark matter candidates and have no $vev$ at $T=0$, the counter-term potential can be written as:
\begin{eqnarray}\label{eq:ct_term}
   V_{\rm CT}(h,\phi_1,..\phi_n,T)   &=& \frac{1}{2} \delta\mu_H^2 h^2 + \frac{1}{4}\delta\lambda_H h^4  +\sum_i^n \frac{1}{2}\delta\mu_{\phi_i}^2 \phi_i^2
\end{eqnarray}
The conditions for calculating the bare coefficients in the above counter-term potential are
\begin{eqnarray}
 \left\{\frac{d}{dh},\frac{d^2}{dh^2},\frac{d}{d\phi_i^2}\right\} \Big(V^{\rm CW}_{1-loop}(h,\phi_1,..\phi_n,T)+V_{\rm CT}(h,\phi_1,..\phi_n,T)\Big) \Big|_{h=v,\phi_i=0,T=0}=0.
 \end{eqnarray}
 Using the above conditions, the coefficients given in Eq.~\eqref{eq:ct_term} can be written as
\begin{eqnarray}
    \delta\mu_H^2&=&-\frac{1}{2v}\left(3 \frac{dV^{\rm CW}}{d h} -v \frac{d^2V^{\rm CW}}{d h^2} \right)\Big|_{h=v,\phi_i=0,T=0},\,\, \delta\mu_{\phi_i}^2=- \frac{d^2V^{\rm CW}}{d \phi_i^2} \Big|_{h=v,\phi_i=0,T=0} \\
    \delta\lambda_H &=& -\frac{1}{2v^3}\left(v \frac{d^2V^{\rm CW}}{d h^2} -\frac{dV^{\rm CW}}{d h} \right)\Big|_{h=v,\phi_i=0,T=0}.
\end{eqnarray}
In the last two equations, we have used the shorthand notation, $V^{\rm CW}\equiv V^{\rm CW}_{1-loop}(h,\phi_1,..\phi_n,T)$. 

\label{Bibliography}
\bibliographystyle{JHEP}
\bibliography{Refs}

\end{document}